%% file: paper.tex
\documentclass[twocolumn,aip,jcp,reprint]{revtex4-1}
\usepackage{amsmath}
\usepackage{amssymb}
\usepackage{graphicx}
\usepackage{time}
\usepackage{siunitx}
\usepackage{pbox}
\usepackage{bm} 
\usepackage{natbib}
\usepackage{natmove}
\usepackage{subfig}
\usepackage{textgreek}
\usepackage[colorlinks,linkcolor=blue,citecolor=blue,urlcolor=blue]{hyperref}
\usepackage{float}
\usepackage{soul}
\usepackage{xcolor}
\usepackage{color}
\graphicspath{ {figures/} }
\begin{document}

\newcommand{\average}[1]{\left\langle {#1} \right\rangle} 
\newcommand{\pd}[2]{\frac{\partial {#1}}{\partial {#2}}}
\newcommand{\cov}{\textrm{Cov}}
\newcommand{\var}{\textrm{Var}}

\title{Configurational Mapping Significantly Increases the Efficiency of Solid-Solid Phase Coexistence Calculations via Molecular Dynamics: Determining the FCC-HCP Coexistence Line of Lennard-Jones Particles}

\author{Natalie P. Schieber}
\affiliation{Department of Chemical and Biological Engineering, University of Colorado Boulder, Boulder, CO 80309, USA}
\author{Michael R. Shirts}
\affiliation{Department of Chemical and Biological Engineering, University of Colorado Boulder, Boulder, CO 80309, USA}

\begin{abstract}
In this study, we incorporate configuration mapping between simulation ensembles into the successive interpolation of multistate reweighting (SIMR) method in order to increase phase space overlap between neighboring simulation ensembles. 
This significantly increases computational efficiency over the original SIMR method in many situations. We use this approach to determine the coexistence curve of FCC-HCP Lennard-Jones spheres using direct molecular dynamics and SIMR.  As previously noted, the coexistence curve is highly sensitive to the treatment of the van der Waals cutoff.  Using a cutoff treatment, the chemical potential difference between phases is moderate, and SIMR quickly finds the phase equilibrium lines with good statistical uncertainty. Using a smoothed cutoff results in nonphysical errors in the phase diagram, while the use of particle mesh Ewald for the dispersion term results in a phase equilibrium curve that is comparable to previous results. The drastically closer free energy surfaces for this case test the limits of this configuration mapping approach to phase diagram prediction.  
\end{abstract}
\maketitle
\newpage
\section{Introduction}

Polymorphism, or the ability of a crystal to pack into multiple metastable states, is important in materials study and design. Polymorphism affects properties of materials such as charge transport~\cite{Stevens2015} and bioavailability~\cite{Chen2009, Bauer2001a}. When multiple metastable polymorphs with different properties are present, the calculation of solid-solid coexistence curves becomes important. Temperature and pressure transformations are present in materials such as pharmaceuticals~\cite{Boldyrev2004, Fabbiani2006}, and metals~\cite{Boehler2000, Choukroun2010}. 

Traditional phase-coexistence calculation methods, such as the Gibbs ensemble method~\cite{Panagiotopoulos1988, Panagiotopoulos1995, Panagiotopoulos2002}, either are not applicable to solid-solid systems, or require a previously known coexistence point and suffer from increasing error due to the use of numerical integration~\cite{Kofke1993, Strachan1999}. We have previously introduced the Successive Interpolation of Multistate Reweighting (SIMR) method to predict solid-solid phase diagrams.~\cite{Schieber2018} This methodology does not rely on lattice dynamics, and thus is applicable in systems that are far from harmonic. It calculates the phase diagram from direct calculation of the relative Gibbs free energy using a series of direct molecular dynamics or Monte Carlo simulations without any specialized sampling techniques. It can thus can be wrapped around any molecular simulation code.

One drawback to this methodology is that it requires an overlap in the energy and volume phase space between adjacent temperature and pressure simulations. This presents a challenge in a number of situations, for example, when an extremely large pressure range is desired. Here, we present an extension of the SIMR method using a configurational mapping technique, inspired by the work of Tan and collaborators~\cite{Tan2010,Schultz2016,Moustafa2015} that reduces the number of simulations required and therefore the computational cost.

We have applied this method to the solid-solid phase diagram of Lennard-Jones spheres, a common test systems in molecular simulation. The Lennard-Jones potential is often used to approximate the solid phase of noble gases such as argon, as well as the highly spherical methane, and to test methodologies for more chemically complex solids~\cite{Hansen1969, Hoover1967, GPollock1976}. 
Many methods have been used to successfully and accurately calculate the melting and vaporization lines of Lennard-Jones system~\cite{Agrawal1995, Smit1992a, Smit1991, Mastny2007, Morris2002, Errington2004, Luo2004, Davidchack2003, Schultz2018}, such as the Gibbs ensemble method~\cite{Smit1992a, Smit1991, Panagiotopoulos1995} (for vapor-liquid) and thermodynamic integration~\cite{Mastny2007}. However, it is significantly harder to calculate solid-solid phase equilibria.

In this paper, we focus on solid infinite crystal systems. While the coexistence line of the hexagonal close packed (HCP) and face centered cubic (FCC) phases of Lennard-Jones spheres has been calculated using a variety of approximations and methods, there is substantial variation in the results from these studies.  All studies agree that the two stable phases of the solid Lennard-Jones spheres are extremely close in free energy, on the range of $\Delta G = 1-10 \times 10^{-4}$ per particle in reduced units throughout most of the phase diagram, meaning that very small uncertainties or errors results in large changes and uncertainties in the phase diagram. 

\begin{center}
\begin{figure}
\centering
{\centerline{\includegraphics[width=0.5\textwidth]{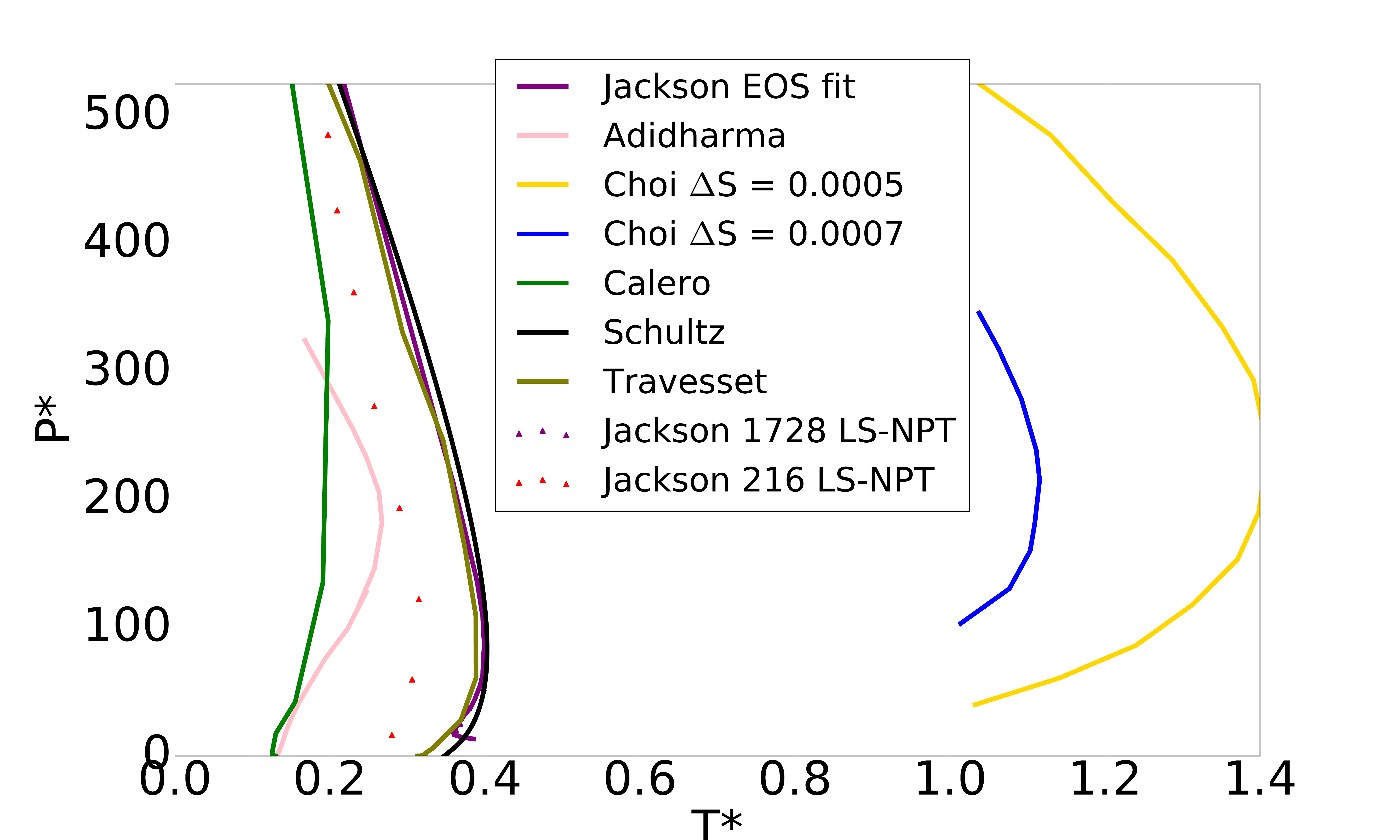}}}
        \caption{\label{fig:prevresults} Previous literature predictions of the FCC-HCP coexistence lines using perturbation theory in Choi et al.~\cite{Choi1993}, Monte Carlo simulations in Adidharma et al., Jackson et al., and Schultz et al.~\cite{Adidharma2016, Jackson2002, Schultz2018}, and dynamics lattice theory with and without anharmonic corrections in Travesset et al.~and Calero et al.~\cite{Travesset2014, Calero2016} show large differences in predicted coexistence regions.}
\end{figure}
\end{center}

A number of research groups have attempted to calculate the Lennard-Jones FCC/HCP phase diagram with a range of approximations, with generally inconsistent results. Choi et al.~\cite{Choi1993} performed an early calculation using perturbation theory of the Lennard-Jones potential around the hard-sphere close packing to determine the phase boundary between the FCC and HCP solids and the liquid phase. Van der Hoef's equations were then used to obtain the residual Helmholtz energy~\cite{vanderhoef2000}. The configurational Helmholtz free energy can then be expanded as a perturbation series and thermodynamic properties can be calculated from the first two terms~\cite{Kim1989, Kang1986}. However, as can be seen in Figure~\ref{fig:prevresults}, this approach was not consistent with later, more comprehensive approaches. 
The coexistence line between FCC and HCP LJ structures has also been calculated using dynamic lattice theory 
(DLT)~\cite{Travesset2014}. However, this approach for obtaining the free energy uses a harmonic approximation, which is not valid at the level of the accuracy needed here.  

Calculation of the fully anharmonic Helmholtz free energy have previously been performed using Monte Carlo simulations~\cite{EPollock1976, Adidharma2016, Jackson2002}. In the work of Adidharma et al.~\cite{Adidharma2016} canonical ensemble Monte Carlo simulations were performed at a variety of reduced temperatures and densities. The results from the simulations were then fit, using the energy and pressure from the simulations, and constants derived by Stillinger~\cite{Stillinger2001}, to the equation for the Helmholtz free energy of the Lennard-Jones solid derived by van der Hoef~\cite{vanderhoef2000}. From the Helmholtz free energy, the coexistence line was then determined. 

Lattice switch Monte Carlo is another approach that has been used to calculate the free energy difference between phases. In the lattice switch Monte Carlo method, a transformation between phases is proposed, which takes the molecules of one structure and converts the atomic positions and box vectors to those of the other phase. A range of multicanonical approaches must be applied in order to get sufficient exchange between the packings~\cite{Bruce1997, Bruce2000}. Once the free energy difference was calculated as a function of $T^*$ and $P^*$, the phase boundary is easily determined. This method has been used to calculate the FCC-HCP coexistence line~\cite{Jackson2002} and study the instability of the BCC phase relative to FCC~\cite{Underwood2015}. The results of Jackson are more consistent with later methods, but are still temperature shifted, especially for smaller box sizes; for larger box sizes, the method was too inefficient to run at higher densities.

One recent comprehensive study of the LJ phase diagram was an extension of the earlier dynamic
lattice theory approach of Travesset~\cite{Travesset2014,Calero2016}. Rather than
directly calculate the full free energy, the anharmonic
contribution to the free energy was calculated using molecular
dynamics using thermodynamic integration along a switching parameter,
$\lambda$, where $\lambda=0$ corresponds to the harmonic potential
energy $U^{DLT}$, and $\lambda=1$ is the full Lennard-Jones potential,
as seen in equation 16 of Calero et al.~\cite{Calero2016}. Using
the DLT energy and 20--50 simulations with a mixed potential $\lambda$
value, the anharmonic contribution is calculated by thermodynamic integration.
Once the harmonic (via DLT) and anharmonic contributions to the
free energy have been calculated, a conversion was performed to
find the corresponding pressure for use in the calculation of the
temperature-pressure coexistence curve. The addition of the anharmonic
free energy term to the previous DLT results illustrates how a small
change in free energy value results in a large change in the
coexistence curve, as seen by the difference in the `Travesset' and
`Calero' lines in Figure~\ref{fig:prevresults}.  However, even more recently, 
Schultz et al.~\cite{Schultz2018} published a comprehensive study of the Lennard-Jones phase diagram, 
using both exhaustively extensive direct simulation and analytic quasiharmonic approaches to 
explore the entire phase diagram of Lennard-Jones particles. Interesting, the results of FCC-HCP equilibria were 
in stronger agreement with Jackson's (partial) results~\cite{Jackson2002} than the later results of Calero et al.~\cite{Calero2016} 

A comparison of many of these previous FCC-HCP coexistence lines over
the full range of methods is shown in Figure~\ref{fig:prevresults} and
shows large differences between methods. All data was taken directly
from temperature-pressure phase diagrams present in the papers using
WebPlotDigitizer~\cite{Rohatgi2018}, with the exception of the Schultz
et al. results, which were taken from correlations lines 7 and 9 of
Table I of Schultz et al.~\citet{Schultz2018}. We note that, as verified with the
authors, the $v^4$ that should be in the denominator of line 9 is incorrectly
written as $v^2$. This was a transcription error in the table
alone, not an error in the results of the study.
However, none of the studies above published explicit error
bars, making comparisons particularly difficult.

\section{Methods}

\subsection{SIMR phase diagram prediction method}

To obtain the phase diagrams of Lennard-Jones spheres using full molecular dynamics, we used the Successive Interpolation of Multistate Reweighting (SIMR) method~\cite{Schieber2018}. This method combines the reduced free energy difference values between temperature and pressure states within a polymorph, defined as $\beta G$, and a reference Gibbs free energy difference between polymorphs at the same temperature, which can then be combined to obtain the Gibbs free energy difference between polymorphs at all temperatures and pressures in the region of interest. 

The reference Gibbs free energy difference value is determined using the pseudo-supercritical path method~\cite{Zhang2012, Eike2006, Eike2005} (PSCP). This method determines the free energy required to take each polymorph from a real crystal to an ideal gas; the free energy between the two polymorphs is then the difference of those values. 

The reduced free energy between states within a polymorph is found using the multistate Bennett acceptance ratio (MBAR)~\cite{Shirts2008}. This method uses equation~\ref{eq:mbar4} to iteratively solve for the reduced free energy of each state $f_{i}$ with respect to each other state $f_{k}$, where $N_{k}$ is the number of configurations drawn from state $k$ and $ u_{k}(x_{jn})$ is the reduced energy of configuration $n$ sampled in state $j$ and evaluated in $k$. The reduced energy is defined as $u_k(x_{jn}) = \beta_k U(x_{nj}) +  P_k V(x_{nj})$. 
\begin{equation} \label{eq:mbar4}
				f_{i} = -  \ln \sum_{j=1}^{K} \sum_{n=1}^{N_{j}} \frac{\exp[- u_{i}(x_{jn})]}{\sum_{k=1}^{K} N_{k} \exp[ f_{k} - u_{k}(x_{jn})]}
\end{equation}
Using the definition of reduced free energy as given above, the Gibbs free energy difference between two polymorphs at state $i$ is then given as equation~\ref{eq:finaldg}, where $\Delta f_{ij}$ is the difference in reduced free energy between states $f_i$ and $f_j$, and $T_{ref}$ is some reference temperature where the $\Delta G_{ij}$ is known.  Linear interpolation is then used to find the points where the difference between polymorphs is zero, which is coexistence. The uncertainty in the coexistence points using this method is found in equation~\ref{eq:uncertainty} where $\delta d$ is the magnitude of the uncertainty in the coexistence line perpendicular to the line, and $\delta \Delta G$ is the uncertainty in the free energy at a point along the coexistence line.  Full details of this method can be found in Schieber et al.~\cite{Schieber2018}. In theory, reweighting can be used to refine estimates in between coexistence points rather than direct interpolation but this is less reliable in the case of configuration mapping, as described below. 

\begin{equation} \label{eq:finaldg}
\begin{split}
			 \Delta G_{ij}(T) = k_B T \Big ( \Delta f_{ij}(T) - \Delta f_{ij}(T_{ref}) \Big ) + \\ 
			\frac{T}{T_{ref}} \Delta G_{ij}(T_{ref}) 	
\end{split}
\end{equation} 

\begin{equation} \label{eq:uncertainty}
\delta d = \sqrt{\left(\frac{\partial \Delta G}{\partial P}\right)^{2}+\left(\frac{\partial \Delta G}{\partial T}\right)^{2}} \delta \Delta G
\end{equation}

\subsection{Configuration mapping}
One requirement for simulations used for the SIMR method is that simulations adjacent in temperature or pressure have a non-negligible amount of phase space overlap~\cite{Schieber2018}, as defined in equation~\ref{eq:overlap4}. Conceptually this means that simulations have some set of configurations that they both sample.  The overlap between states 1 and 2 is then dependent on the probability of all configurations, $x$, in each of the two distributions, $P_{1}$ and $P_{2}$. Due to this requirement, the number of simulations performed is dependent on the width of the energy and volume distributions of the simulations. Systems with wider potential energy and volume distributions are likely to still achieve phase space overlap with wider spacing. The width of these distributions, and thus the spacing in temperature and pressure that is allowable between simulations, depends on factors such as the temperature, pressure, and the size and flexibility of the molecule. 
In order to decrease the number of simulations and therefore the computational resources required, it is desirable to increase the spacing between sampled states by increasing phase space overlap between states $O_{1,2}$. 
\begin{equation} \label{eq:overlap4}
O_{1,2} = \int_{x \in \Gamma} \frac{P_{1}(x) P_{2}(x)}{P_{1}(x)+P_{2}(x)} \,dx
\end{equation}
One potential way to increase phase space overlap between states is configuration mapping~\cite{Tan2010,Tan2010e,Schultz2016,Moustafa2015,Paliwal2013, Jar2002}. Configuration mapping transforms the set of coordinates in one thermodynamic state into a set of coordinates that is more likely to have a low energy in the the other thermodynamic state of interest, and evaluate the energy in the new state with the transformed configuration rather than the originally sampled configuration.  We can then analytically calculate the free energy change for performing this mapping. This approach was used by Tan et al.~\cite{Tan2010,Tan2010e} to calculate the temperature dependence of the free energy of solids, by Paliwal et al.~\cite{Paliwal2013} to calculate the Gibbs free energy of transformation between different water models~\cite{Paliwal2013}, as well as in a differential form to calculate physical properties such as the heat capacity of HCP iron and the dielectric constant of the Stockmayer potential.~\cite{Schultz2016} Configuration mapping was shown to significantly improve precision of these calculations. Here, we propose to use this methodology to improve the phase space overlap for SIMR. 

Mathematically, we define a transformation $T(x)$, with Jacobian
$J(x)$, which is allowed to depend on the current configuration $x$,
and $\Delta U(x)$ is the difference in potential energy between the
configuration when evaluated in the original and mapped
states~\cite{Moustafa2015}, $\Delta U(x) = U(T(x)) - U(x)$. In general, the potential can also change~\cite{Paliwal2013}, but in this study, we only change the temperature and pressure between states.

For a simple one-step transformation using the Zwanzig equation, the Helmholtz free energy difference in terms of the mapping can be written as:
\begin{equation} \label{eq:mapping}
\Delta A = -k_B T \ln\langle |J(x)| e^{-\beta \Delta U(T(x))} \rangle
\end{equation}
More generally, when using configuration mapping to calculate free energy differences with multistate reweighting, we can derive equivalent formulas by replacing the reduced energy $u(x)=\beta U(x) + \beta PV$ used in MBAR~\cite{Shirts2008} or BAR~\cite{Bennett1976} in the NPT ensemble with a ``warped'' reduced energy, defined in Eq.~\ref{eq:warped} by analogy with ``warped bridge sampling'', a version of this technique used in statistics~\cite{Meng2002} to calculate the free energy difference between states.  In Eq.~\ref{eq:warped}, $i$ is the state the configuration was drawn from, $j$ is the target state the energy is evaluated in, $T_{ij}(x)$ is the transformed set of coordinates that were sampled from state $i$, and $|J_{ij}(x)|$ is the determinant of the Jacobian of the transformation $T_{ij}$. 
\begin{equation} \label{eq:warped}
u_{ij}^{w} = u_{j}\left(T_{ij}(x_{i})\right) - \ln |J_{ij}(x_{j})|
\end{equation}
Equation~\ref{eq:warped} is applicable for all transformations that have a nonsingular Jacobian, but only a relatively small proportion transformations actually increase phase space overlap. A good transformation is one that is both relatively easy to implement and results in significant increase in overlap. With a good transformation, the efficiency of the free energy calculation from state $i$ to $j$ can be made much more efficient.

In this study, we applied this coordinate mapping approach to the system of Lennard-Jones spheres to map between states defined by different temperature and pressures. In the case of point particles such as Lennard-Jones spheres, we only need to map the location of the particles themselves, and not deal with any internal degrees of freedom. This scaling is applied between any two pairs of simulations, as the average box vectors change both through thermal expansion and compression/expansion due to changes in pressure.

To implement this type of transformation, first, the trajectories from states $i$ and $j$ are read.  The desired transformation between the two states is determined, which usually requires some information from both trajectories to be efficient.  This transformation is then applied to the coordinates in the trajectory $i$. In our case, these new coordinates are written to a new trajectory file and the energy of each frame of the trajectory is then reevaluated. The warped reduced energy is calculated using the warped energy as well as the new $P$ and $T$, and the reduced free energy is calculated using equation~\ref{eq:mbar4}. For the multistate reweighting process used in SIMR, this transformation/reevaluation is performed for every set of pairs states in the (not necessarily regular) $T,P$ grid.

The transformation we used is defined by:
\begin{eqnarray} \label{eq:ljmap}
\vec{r}_{i,r} &=& \vec{r}_{i} B_{i}^{-1} \nonumber \\ 
\Delta \vec{r}_{i,r}^{w} &=& \Delta \vec{r}_{i,r} \left(\frac{T_{j}}{T_{i}}\right)^{1/2} \nonumber \\ 
\vec{r}_{i,r}^{w} &=& \vec{r}_{i,r} + \left(\Delta \vec{r}_{i,r}^{w} - \Delta \vec{r}_{i,r}\right) \nonumber \\
\vec{r}_{i}^{w} &=& \vec{r}_{i,r}B_{j}
\end{eqnarray}
where $\vec{r}_{i}^{w}$ is the new coordinate in the target ensemble, $\vec{r}_{i}$ is the original coordinate, and $B_{i}$ and $B_{j}$ are the average box vector matrices of the original and target trajectories.
If two simulations differ in temperature, we also scale the deviation of the particle from its equilibrium position as derived by~\citet{Tan2010e}. This temperature scaling is carried out using the reduced coordinates $\vec{r}_{i,r}$, or the fractional coordinates within the box, thus making this a three step process.  In Eq.~\ref{eq:ljmap}, $\Delta \vec{r}_{i}$ is the magnitude of the deviation of the molecule from its equilibrium position, $\Delta \vec{r}_{i} = \vec{r}_{i} - \langle \vec{r}_i\rangle$. and $T_{i}$ and $T_{j}$ are the temperatures of the initial and target trajectories. The energy contribution of this Jacobian is $\frac{3N-3}{2}\log
\frac{T_j}{T_i} + 3N\log \frac{|B_j|}{|B_i|}$, where $N$ is the number of
particles. The `-3' occurs because of the removal of translational center of
mass motion in the simulation. Note in this case a constant Jacobian
is used for all configurations, though in theory it can be configuration-dependent.  Once all of the $u_{ij}^{w}$ values have
been calculated, eq.~\ref{eq:mbar4} can be used directly to calculate the reduced
free energy differences between states.

\begin{center}
\begin{figure}
\centering
\subfloat[]{\centerline{\includegraphics[width=0.5\textwidth]{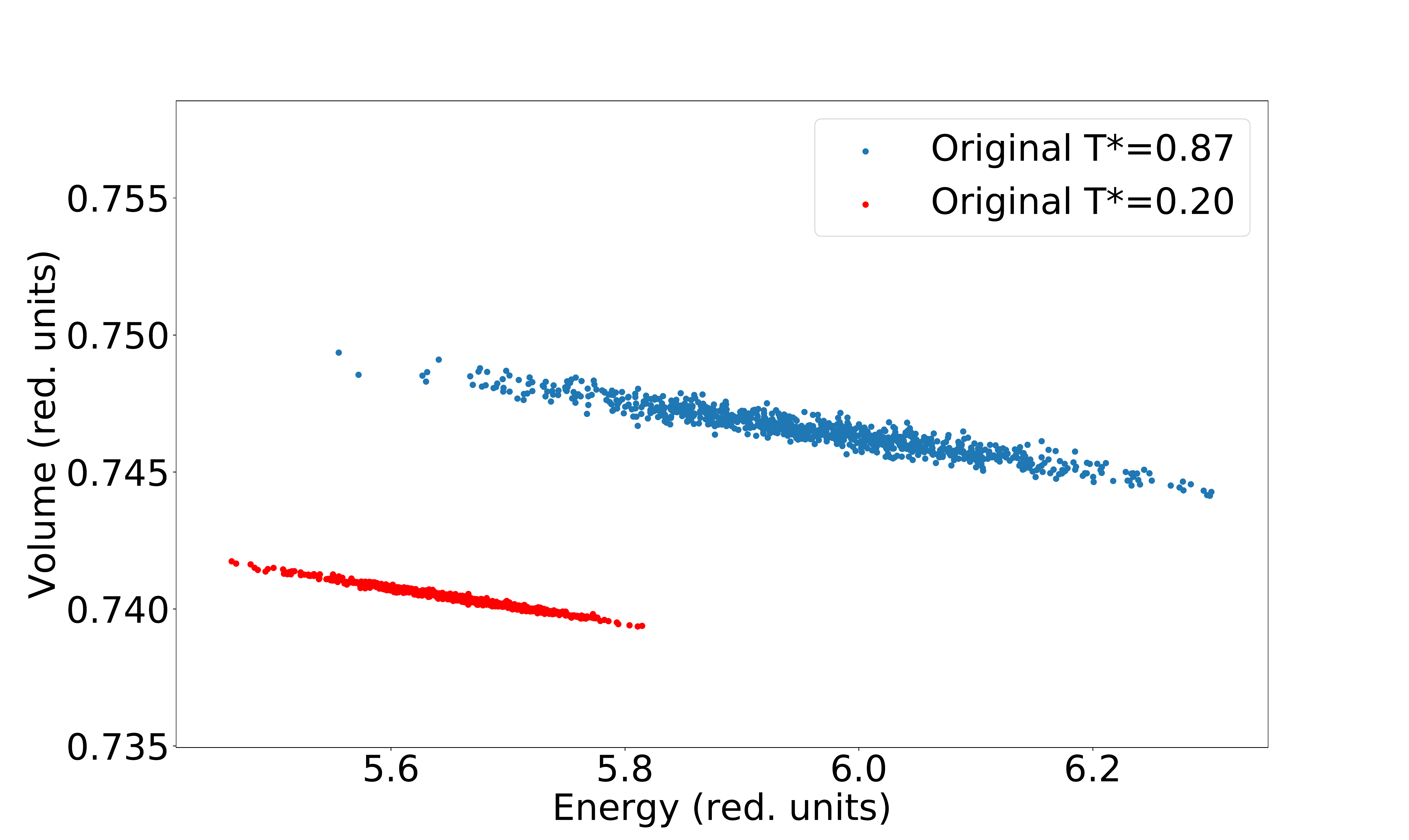}}}
\newline
\subfloat[]{\centerline{\includegraphics[width=0.5\textwidth]{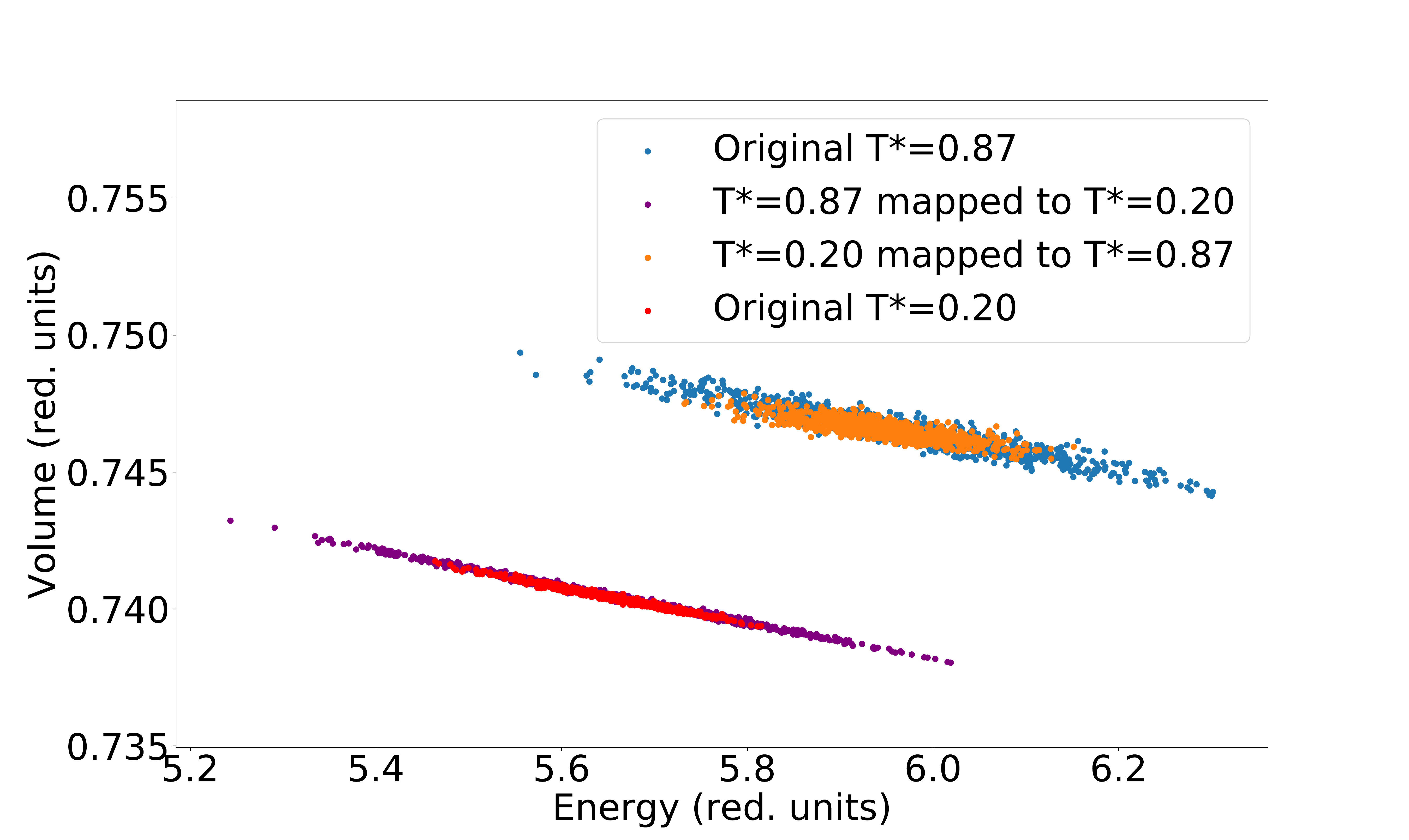}}}
\newline
        \caption{\label{fig:ljpdoverlap} Energy and volume distributions of two trajectories of Lennard-Jones spheres carried out at different temperatures, at $152 P^*$ with both the original trajectories (a) and original and mapped trajectories (b), display drastically increased phase space overlap using configuration mapping as compared to the original trajectories.}
\end{figure}
\end{center}

An example of the effect of this mapping on the energy-volume distribution for two temperature and pressure states in this LJ system can be seen in Figure~\ref{fig:ljpdoverlap}. This figure shows the difference between the overlap in energy and volume achieved between two states unmapped and using configuration mapping. The two unmapped trajectories show no overlap of their energy and volume distributions. The mapped distributions, however, shows significant overlap in energy and density, which in most cases will translate directly to overlap of configuration phase space. 

The cost of energy reevaluations is very low compared to the cost of simulations. For example, in one test on a standard laptop with GROMACS, the cost of mapping and reevaluating uncorrelated samples between two states from a 9 ns simulation costs approximately $7.4$ CPU-min. The cost of running one LJ particle mesh Ewald (PME) simulation itself for 9 ns is approximately is 56 CPU-hrs. For a set of 187 states, as is examined here, the mapping cost between all pairs of states is then $\sim 4300$ CPU-hrs, which is the cost of about 75--80 additional simulations. However, by using mapping, we need at least 15--40$\times$ less simulation than we otherwise would have needed with SIMR, as discussed below, though the exact numbers will depend significantly on the specific code used.

When this particular mapping was applied to the system of
Lennard-Jones spheres, the number of states required to achieve
overlap decreased significantly. Without mapping, the minimum pressure
spacing required to achieve sufficient overlap for the calculation of
MBAR to converge with a finite value was approximately 1 $P^*$. With
mapping, this could be increased to 25 $P^*$ for the cutoff-based
Lennard-Jones simulations and $12.5 P^*$ for the higher precision PME
calculations, which translates directly into an efficiency increase of
12--25 by decreasing the number of simulations required. Although
overlap in the temperature dimension was already enough to achieve MBAR convergence with
simulations every $0.066 T^*$ ($0.033 T^*$ for PME simulations),
mapping decreased uncertainty of free energies between neighboring
states in the $T$ direction by roughly a factor of 1.31 to 1.48 (for example,
at $P^*=127$ with simulations spaced at $T^* =0.006$ or $T^* = 0.026$ respectively), leading to an additional efficiency improvement of
between $~\sim 1.31^2 \approx 1.71$ to  $~\sim 1.48^2 \approx 2.19$, for an overall efficiency gain of 15--40$\times$.

\subsection{Simulation details}

The Lennard-Jones phase diagram was produced using a system of 1200 LJ spheres and the standard Lennard-Jones 12-6 potential. The systems were set up with 10 layers of atoms in the $x$ and $y$ directions and 12 layers of atoms in the $z$ direction, for a total of 300 FCC unit cells and 200 HCP unit cells. 
In a limited study of Jackson et al.~\cite{Jackson2002}. a change of size from 216 to 1728 gave rise to a shift of about 0.1 $T^*$ in the FCC-HCP coexistence line. Since our study is nearer the upper end, size dependence will be relatively small compared to the uncertainty, as discussed in more detail later when we look at possible reasons for differences from previous results. 
The Lennard-Jones parameters for OPLS-UA methane were used in the simulations themselves ($\sigma = 0.373$ nm, $\epsilon = 1.2304$ kJ/mol, m=16.043 amu)~\cite{Jorgensen1984}  though all results are reported in reduced units. The range of the phase diagram was chosen to correlate with previous lattice dynamics studies of Lennard-Jones spheres~\cite{Travesset2014}. The temperature in reduced units was 0.066 to 0.466 and the pressure was between 0.003 and 508.9, which corresponded to 10 to 70K and 1 to 200001 bar in real units. This temperature range was chosen in order to include the region of predicted coexistence without including the region of melting. The pressure range was chosen to include the maximum HCP temperature stability point and the reentrant behavior and to include high enough pressures that the coexistence line is approximately linear and can be extrapolated. Simulations were initially spaced every 0.066 $T^*$ and 25.44 $P^*$. For PME simulations, simulations were spaced every 0.033 $T^*$ and $12.72 P^*$. The largest current limitation in the number of states which can be simulated is the limit on the memory which is available to run MBAR, which will fail with a memory error in the current implementation of \texttt{pymbar}~\cite{Shirts2008} used to solve if the input matrix is too large.

All production molecular dynamics simulations of Lennard-Jones spheres were performed with GROMACS 5.1.2~\cite{Berendsen1995, Abraham2015}, using a velocity Verlet integrator and Nos\'{e}-Hoover temperature control~\cite{Evans1985} with a time constant of 1.485 reduced units (2 ps). Isotropic Martyna-Tobias-Tuckerman-Klein (MTTK)~\cite{Martyna1996} pressure control with a time constant of 7.24 reduced units (10 ps) was used.  This combination of integrator and pressure control was chosen because it was shown to be the most stable for NPT simulation of small LJ systems in GROMACS at high pressure. For PSCP simulations, the \texttt{sd} integrator (Langevin dynamics) and thermostat and the Parrinello-Rahman barostat~\cite{Parrinello1981}
 were used to avoid nonergodicities at the fully restrained and non-interacting states of the PSCP. All particle mesh Ewald (PME) simulations were run for 9 million steps at a time step length of 0.00297 $t^*$ (4 fs) for a total simulation time of 26,728 $t^*$ The cutoff used in these PME simulations was 3.5 $\sigma$. Potential switch simulations were run with the same time step for 11,879 $t^*$. Two types of simulations were run to determine the phase diagram: with cutoffs, and using particle mesh Ewald (PME).
For cutoff-based simulations, the van der Waals interactions were treated with a potential switch cutoff of 1.119 or 0.9325 nm, which corresponds to $3.0 \sigma$ and $2.5 \sigma$ respectively (in units of Lennard-Jones radius). This method smoothly switched the potential over a range of 0.02 nm as a function of radius down to make the potential at the cutoff $0$ using the vdw-modifier \texttt{potential-switch} keyword. Because of the smaller relative energy difference between phases, all PME simulations were run for a factor of 2.5$\times$ longer than all simulations with the potential cutoff to add statistical accuracy.

Rather than increasing the cutoff size and extrapolating to infinity, we used particle mesh Ewald for dispersion interactions to incorporate long-range effects. This method works by only calculating the short range interactions directly. Long range interactions are calculated with a 3D fast Fourier transform on a grid.~\cite{Essmann1995, Petersen1995} The smooth version of this method, as implemented by Essmann et al.~\cite{Essmann1995} uses a B-spline interpolation on a grid to increase computational efficiency.  This method was implemented for the dispersion term in GROMACS by Wennberg et al.~\cite{Wennberg2013, Wennberg2015}.  For all PME simulations, a Fourier grid of $n_x = 36$, $n_y=32$, and $n_z=36$
and a PME grid interpolation order of 6 were used. A cutoff of the direct-space summation was used, and shifted from $3.47319 \sigma$ (1.2955 nm) to $3.5 \sigma$ (1.3055 nm) 
to guarantee smooth integration of the equations of motion, with tolerance of the direct space error at the cutoff (\texttt{ewald-rtol-lj}) equal to $1.0 \times 10^{-6}$. 

Two PSCP calculations were carried out for each calculation, one at 127 $P^*$ and 0.33 $T^*$ and one at 152 $P^*$ and 0.4 $T^*$ in the NVT ensemble with a PV correction term, of $P \Delta \langle V \rangle$  to convert between Helmholtz and Gibbs free energy. Simulations were carried out in NVT at the average volume corresponding to the equilibrated corresponding NPT simulation. The (127 $P^*$, 0.33 $T^*$) phase point was used for the phase diagram calculation, with the other point used to check cycle closure. 
In the PSCP, intermolecular interactions were turned off quartically, while simultaneously harmonic restraints were added to the average lattice positions. A set of 25 intermediates was used, with the force constant turned on quadratically to a maximum value of 113.076 in reduced units ($1000~\mathrm{kJ}\cdot\mathrm{mol}^{-1}\cdot \mathrm{nm}^{-2}$), as per the protocol of Dybeck et al.~\cite{Dybeck2016}. 

Our more challenging phase diagram, generated using PME, is the result of one set of additional low pressure and temperature simulations, which doubled the number of simulations in the region below $200 P^*$ and $0.3 T^*$, after the initial simulations. In this case, unsimulated states, where the average box vectors and displacement vectors are not known, cannot be incorporated into MBAR because the mapping procedure requires knowledge of the equilibrium P and T at each state, which must be obtained from simulation. To calculate the uncertainty, 200 bootstrap samples of simulation configurations were generated. The uncertainty in the $\Delta G$ per particle between the FCC and HCP phases at each point was then determined from the standard deviation of the reduced free energy using the bootstrapped input. This was then converted to the uncertainty in the width of the coexistence line and plotted perpendicular to the line, as errors along the line do not actually affect the line, as described in Schieber et al.~\cite{Schieber2018}. Alternately, using each of the bootstrapped free energies, a new phase equilibrium line for each bootstrap resampling line can be drawn. Error bounds can then be generated by taking the lines that represent a given confidence interval away from the median line. This is most accurately done in the perpendicular direction, though because in this case the phase line is mostly vertical and noise in the line makes tangent determination challenging, we look at the confidence interval in the temperature dimension. We use a linear spline for the phase equilibrium points obtained using SIMR to sort the points at any temperature of interest. A comparison of the two types of error analysis can be found in section II of the supplementary material (using a $1\sigma$ confidence interval). We note very good consistency between the methods above the $100 P^*$ line; below the line the bootstrap estimate becomes inconsistent because of low overlap, leading to large fluctuations in the bootstrap uncertainty. 

Thermodynamic cycle closure was used to validate the process for the PME line, using two separate PSCP calculations at two different points. At 127 $P^*$ and 0.40 $T^*$ the PSCP value is $-0.000898(9)$. At 127 $P^*$ and 0.27 $T^*$ the PSCP value is $0.000166(9)$. Using the second PSCP at the reference value in SIMR, the calculated $\Delta G$ at 127 $P^*$ and 0.40 $T^*$ is $ -0.00086(6) $, which is within uncertainty of the PSCP value at that point, indicating cycle closure to high precision. After initial simulations and one round of additional simulations at areas of low overlap, at $T^* = 0.20$ the average uncertainty in $\Delta\mu$ was $0.0004$ reduced units is similar to that shown in Figure 6 of Calero et al.~\cite{Calero2016}, which ranges from approximately $0.00025$ to $0.0006$, as seen in our Figure~\ref{fig:dgvp}.  All free energies were calculated using the \texttt{pymbar 3.0.3} implementation of MBAR~\cite{Shirts2008}.

\section{Results}
\subsection{Molecular dynamics phase diagrams of Lennard-Jones spheres}

\begin{center}
\begin{figure}
\centering
{\centerline{\includegraphics[width=0.5\textwidth]{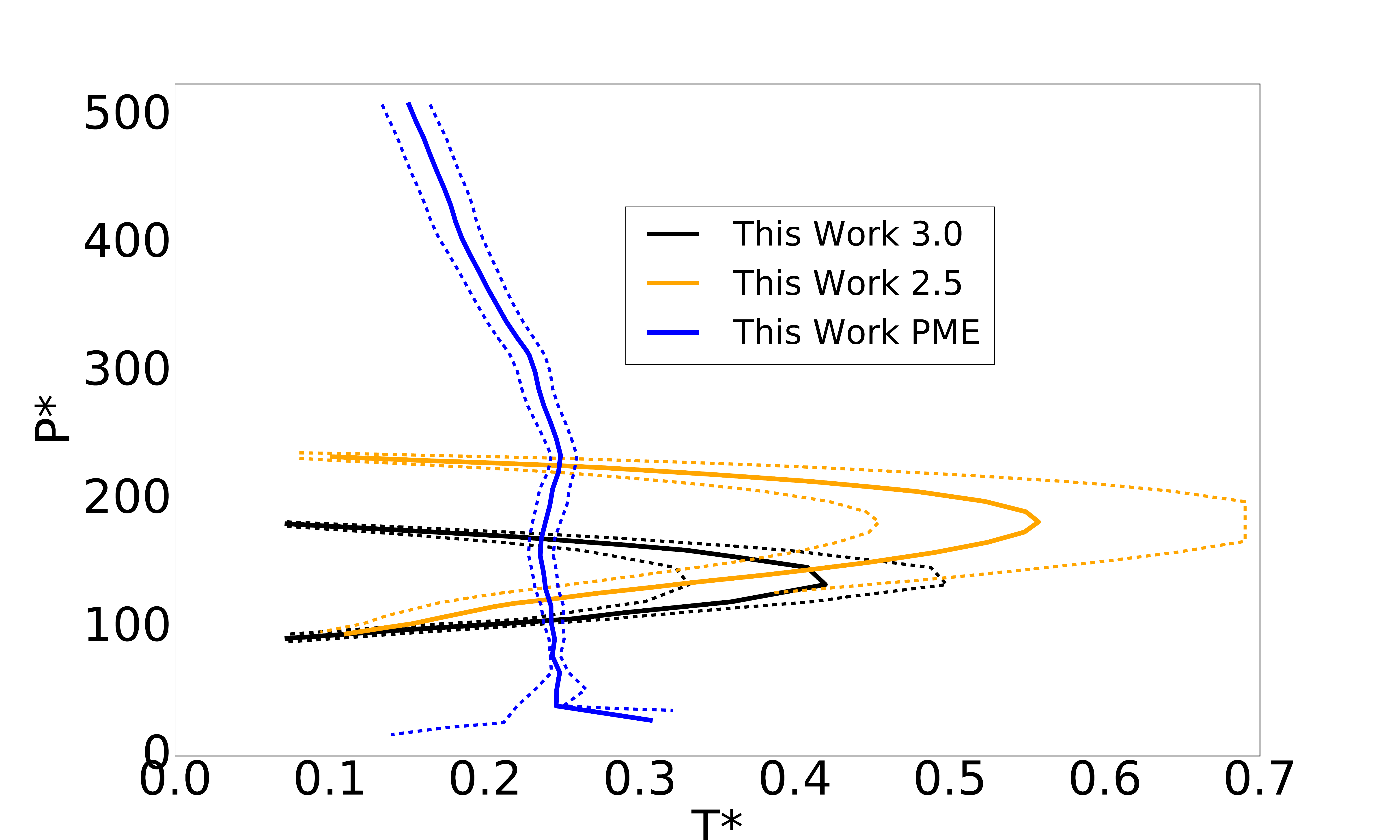}}}
        \caption{\label{fig:ljsimr} The phase diagram of Lennard-Jones spheres using the SIMR method and a potential switch cutoff shows higher pressure stability when using a $2.5 \sigma$ cutoff than a $3.0 \sigma $ cutoff and the highest pressure stability when using PME treatment of long range interactions. The PME results used substantially more simulations, as the FCC and HCP free energies are much closer to parallel when PME is used, meaning lower uncertainties are much harder to obtain. Uncertainty on the SIMR results represented by dashed lines.}
\end{figure}
\end{center}

\begin{center}
\begin{figure}[h]
\centering
{\centerline{\includegraphics[width=0.5\textwidth]{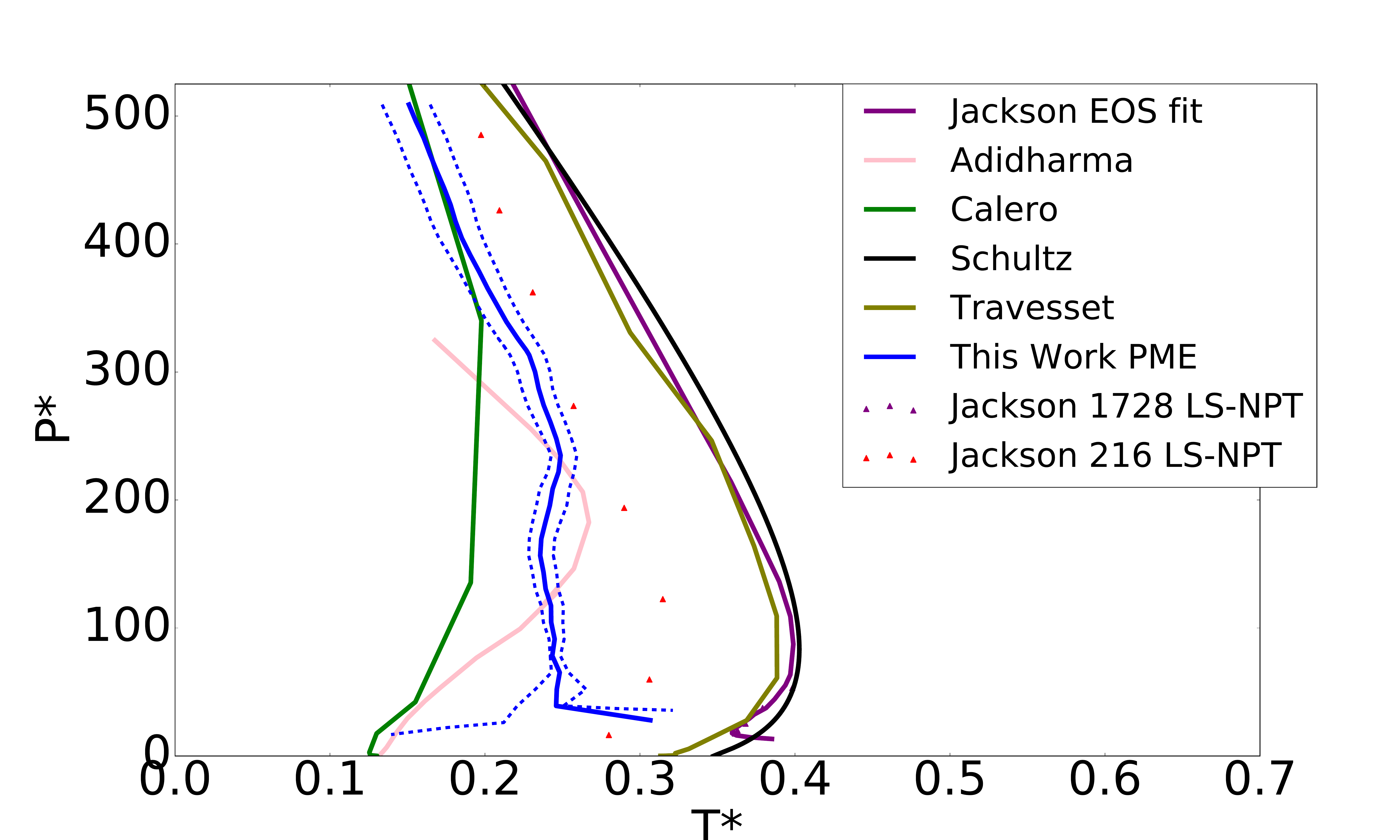}}}
        \caption{\label{fig:ljcompcutoff} The phase diagram of Lennard-Jones spheres using SIMR (labeled `this work PME') plotted with a number of literature results, is close to consistency within uncertainty of previous higher-precision results~\cite{Calero2016}. Uncertainty on the SIMR results represented by dashed lines}.
\end{figure}
\end{center}

We see in Figure~\ref{fig:ljcompcutoff} that using mapping, the coexistence line was determined with good precision. The phase coexistence line is significantly affected by the treatment of the cutoff. Cutoffs should not affect the liquid-vapor transition significantly, since those phases are essentially uncorrelated outside of the cutoffs investigated here. However, in solids, cutoff effects of calculations using LJ spheres are important~\cite{Calero2016,Trokhymchuk1999}. 
The coexistence line predicted by the SIMR and configuration mapping using a $2.5 \sigma$ cutoff has a region of HCP coexistence that 
is higher in pressure and temperature, and spans a wider range of pressures than the coexistence line predicted using a $3.0 \sigma$ cutoff, as seen in Figure~\ref{fig:ljsimr}. Both of these results are a poor match for literature results extrapolated to long cutoff in the high pressure region, as seen in Figure~\ref{fig:ljcompcutoff}.  
The reentrant behavior resulting from the SIMR method is much sharper than literature results with extrapolated large cutoffs and the high pressure and low temperature region is not well approximated by this method. The poor match of the coexistence lines using a potential switch cutoff is due to the uneven increase in the contribution of the dispersion energy as the pressure is increased. The sudden inclusion of an entire shell of atoms under the cutoff value as pressure increases causes nonphysical behavior in the energy difference between phases. This is a known effect, analyzed for example by Jackson et al.~\cite{Jackson2002}, and we analyze the reasons and resulting issues in more detail in Section I of the supplementary material.

By using particle mesh Ewald for the dispersion term, we obtain a more accurate molecular dynamics phase diagram of FCC and HCP LJ spheres without the inconsistencies introduced by a potential cutoff. This phase diagram shows the same stability trends and reentrant behavior as literature results which in theory include the full statistical mechanics. The resulting PME SIMR phase diagram is approximately a standard deviation away from the results of Calero et al.~\cite{Calero2016}, which at $P^*=250$ is an average of $0.0234 T^*$, and is within uncertainty for most of the pressure range. The maximum HCP temperature stability is somewhat higher than their results. We note that although Calero et al.~do not plot uncertainties in their phase diagram, their Figure 6 includes free energy differences between phases along the $P^*=290$ isobar; this can be used to back-calculate an uncertainty of about 0.01--0.015 in $T^*$ over this part of the range, slightly smaller than ours. The HCP temperature in our simulations stability is, however, lower than the results of Schultz et al.~\cite{Schultz2018}

There are a number of factors that could potentially explain the discrepancies between other results that are 
statistically mechanically complete and our results; however, we find that most of them do not affect the phase diagram.
The major methodological difference between
previous anharmonic results and our PME SIMR results is the treatment
of the cutoff in particular our use of PME to account for long range
interactions, rather than extrapolation to infinite cutoff.  The use
of PME and the parameters used for PSCP were validated to the results of Stillinger et al. \cite{Stillinger2001}, with which multiple groups have gotten highly consistent results.~\cite{Schultz2018}. Our PME parameters gave energies at the lattice minimum of 
FCC and HCP lattice minima within of $0.0001 E^*$ of the Stillinger results, and $\Delta E^*$ between the phases of $7.3 \times 10^{-7} E^*$. Changing the PME cutoff
parameters did not change the average potential energy to well within
uncertainty. The difference in energy between HCP and FCC at the lattice minimum
was $\Delta E^* = -8.70 \times 10^{-4}$.  Increasing the Ewald direct space cutoff by
0.25 $\sigma$ changed this energy difference by $8.5 \times 10^{-6}$, less
than 1\% of the total relative energy, and significantly below the
statistical uncertainty.  Increasing the number of Fourier points by
50\% changed the energy by $1.34 \times 10^{-5}$, approximately
$1.5\%$ relative error.  Additionally, these terms cancel --- when both
increases in accuracy of the calculation are used, the total change in
energy is $3.4 \times 10^{-7}$, less than 0.5\% relative error. This degree
of relative bias in the energy calculation should be roughly
invariant, or even decrease, throughout the simulation, as the energy
becomes dominated by the repulsive term that is entirely short range
at higher pressures. Using a 1200 atom system and the PME settings described above, the initial structures were
minimized for volumes ranging from $182.58 \mathrm{nm}^3$ to $547.71 \mathrm{nm}^3$.
The minimum enthalpies were then compared against the values in Stillinger~\cite{Stillinger2001}, 
and the coexistence pressure was found. 
The $T^*=0$ coexistence volume agreed to within $1.50 \times 10^{-4} V^* $ and the enthalpy at that coexistence point agreed to within $0.136 E^*$, which is within 0.2\%. Our results were also statistically tested for consistency by extrapolating the coexistence line at high pressures (where it is roughly linear) down to the 0 K point, though we did not simulate in this high pressure region. We found that our extrapolated coexistence line crosses the $T^*=0$ line between $P^*=845$ and $P^* = 958$, which is within uncertainty of the literature results of $P^* = 878$.

There are several other system differences which could in theory be
causing the lower temperature coexistence in our results.  One such
effect is system size. Schultz et al. found
that the system size effects are almost entirely harmonic.~\cite{Schultz2018} To study
these effects, we added a harmonic $\Delta G$ correction, equal to the difference in the harmonic free energy of
each phase between a system of 1200 atoms and a system of infinite
cutoff and size, with data graciously provided by the authors of Schultz et al.~\cite{Schultz2018}. However, 
this correction term only shifted the phase diagram up in temperature by an average value of $0.002 T^*$.

The effect of system size and cutoff was also tested at the $T^*=0$ K
line. A series of PME cutoffs between $3.2 \sigma$ and $5.3 \sigma$
were tested, and the $T^*=0$ coexistence crossing was identified. 
The coexistence pressure was shifted by a value of $5
P^*$ between the smallest and largest cutoffs; using the cutoff value of $3.2 \sigma$ the coexistence
pressure is $878.26 P^*$ and at $3.6 \sigma$ the pressure is $883.26
P^*$.
Changing the system
size from 1200 to 9600 atoms produced a change in energy at
coexistence of $7.27 \times 10^{-6} E^*$ and moving from the smallest
to largest cutoff produced a change of $7.61 \times 10^{-6} E^*$.
Moving from a system size of 1200 to 9600 atoms produced a shift in
coexistence of $7 P^*$, which we considered to be sufficiently
accurate, given that our main focus is the $T^*>0$ portion of of the
phase boundary. 

Another difference from the result of Schultz et al. is that they
considered anisotropic expansion of the HCP box, which slightly shifts
the phase coexistence curve in the direction of the HCP
phase. However, their results (in Table III of Schultz et al.~\cite{Schultz2018}) show
that this effect changes the location of the FCC-HCP-vapor triple point by $\Delta
T^*=0.0004$, and the location of the $T^*=0$ intersection of the
coexistence curve by $\Delta P^*=0.01$, and thus is several orders of
magnitude below the other differences considered in this paper.

After analyzing these other effects, we find that the maximum HCP stability
temperature is almost entirely determined by the reference value
obtained from the PSCP. If the PSCP reference value was shifted down by
$0.0007$ $E^*$, our results would be within uncertainty of the results
of Schultz et al, with our maximum HCP stability raised to
$T^* = 0.402$ compared to $0.40$ in that study.
Alternatively, if we use the coexistence point of Schultz et al.\cite{Schultz2018} of $P^*=127, T^*=0.4$ (obtained using the correlations found in Table I), as a reference, instead of the PSCP calculation, the maximum HCP stability moves to $0.42(2)$, and the $T^*=0$ coexistence line still 
intersects between 853 and 1238 $P^*$, which are both within uncertainty of literature results. The $P^*=0$ intersection could not be calculated due to poor overlap in the low pressure region with the simulated data set.
Though this error in the reference $\Delta G$ generated using PSCP is small, it is several times the uncertainty in the calculation, indicating some source of bias exists.
This bias could be caused by several factors, including the lack of anisotropy in the PSCP simulations and finite size effects in that calculation, or very minor errors in the Parrinello-Rahman barostat implemented in GROMACS, which have been noted earlier~\cite{Shirts2013}. We emphasize that any potential errors are so small so as to likely only be noticeable in calculations of this precision. 

Other methods could be used to generate the PSCP, for example, lattice-switch Monte Carlo at a single $(T^*,P^*)$ point, though the PSCP method is the most general for arbitrary crystal packings. In fact, the moves used in lattice-switch Monte Carlo themselves define a configuration mapping between the two phases, and the free energy between two phases can be determined directly from two simulations using Eq.~\ref{eq:warped}, without ever actually implementing lattice switch Monte Carlo. However, preliminary testing demonstrated that this approach was far too inaccurate for the LJ phase diagram because of the low overlap between the mapped FCC$\rightarrow$ HCP and HCP$\rightarrow$FCC ensembles. This low overlap is not surprising given that lattice switch Monte Carlo requires additional acceleration methods to yield reasonable results~\cite{Jackson2002}. 

One challenge with studying the FCC and HCP structures using PME is the extremely small free energy differences between the polymorphs, as seen in Figure~\ref{fig:dgvp}, almost two orders of magnitude lower than the cutoff simulations. These small energy differences mean that longer ($2.5\times$ the potential switch simulations) and more closely spaced simulations are required to obtain sufficiently precise results. At pressures below $100 P^*$, the difference in free energy was small enough that sufficient precision to clearly resolve the phase diagram could not be achieved. This was due to consistently poor phase space overlap at these states, where the larger amount of movement in the atoms causes mapping to work poorly.


\begin{center}
\begin{figure}[h]
\subfloat[]{\includegraphics[width=0.5\textwidth]{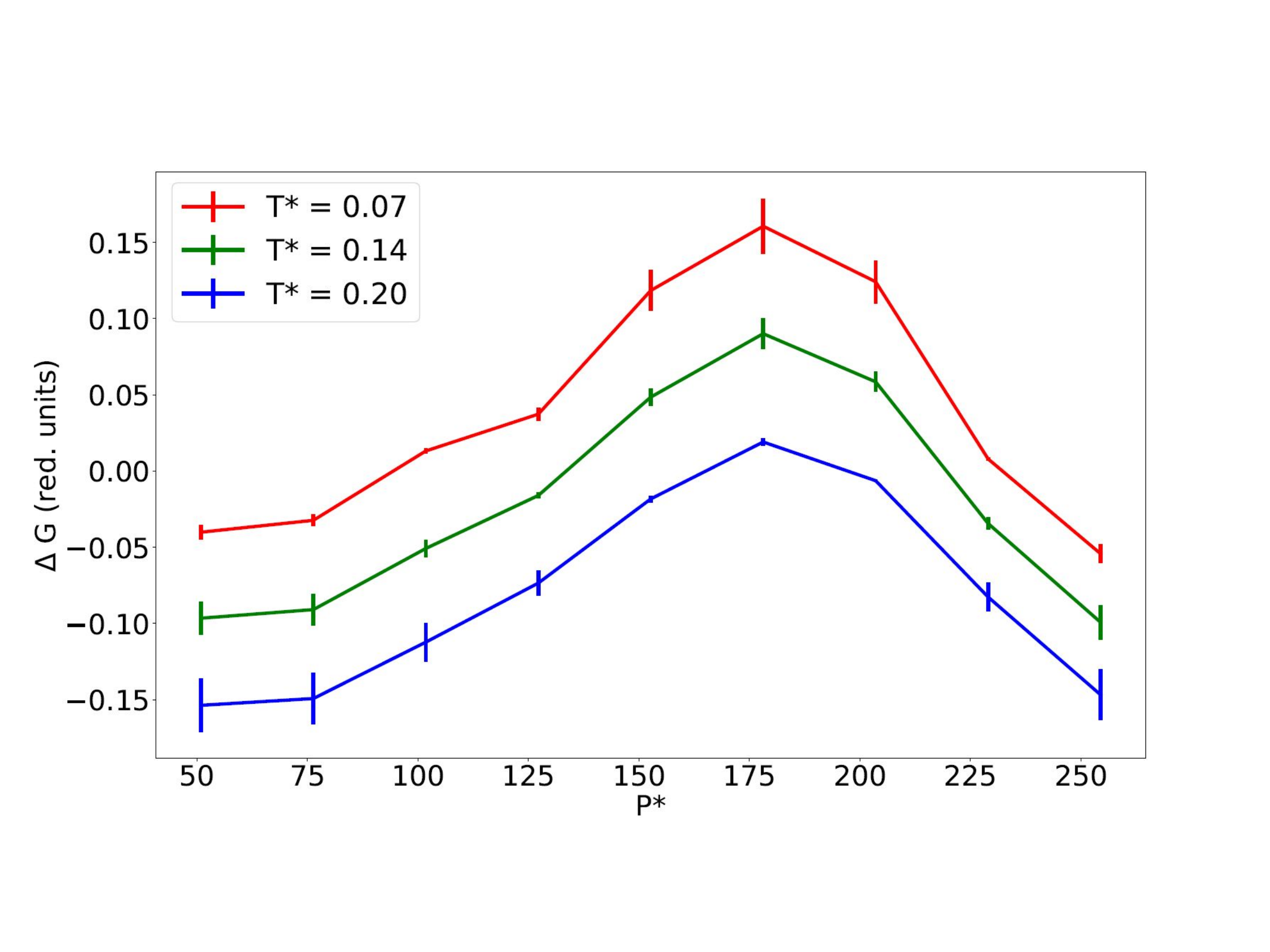}}
\subfloat[]{\includegraphics[width=0.5\textwidth]{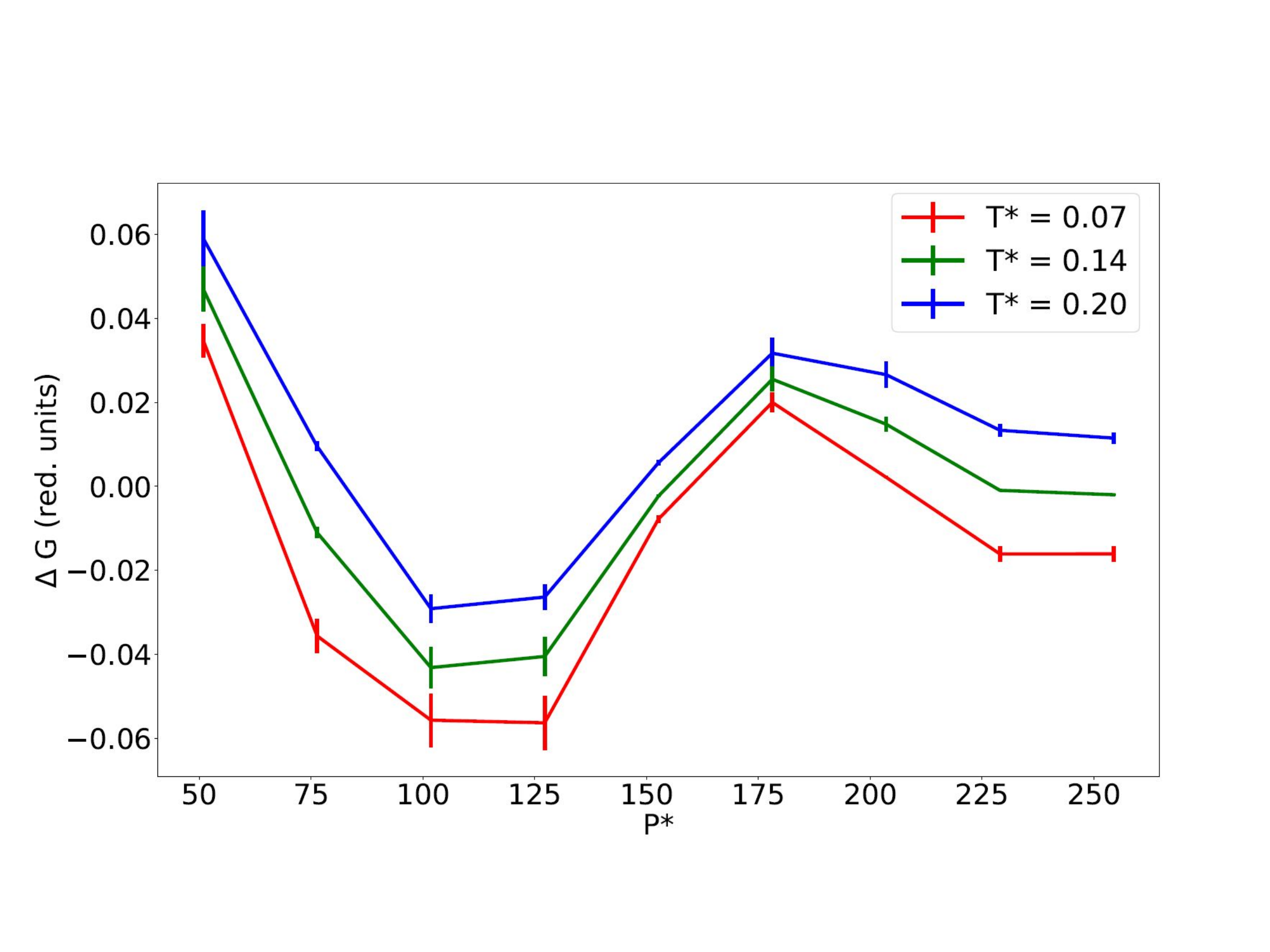}}
        \caption{\label{fig:dGvPcutoff} The FCC-HCP free energy difference per particle and the associated uncertainty as a function of $P^{*}$ for the calculations using a potential switch treatment with cutoffs at (a) $2.5\sigma$ (top) and (b) $3.0 \sigma$ for the long-range electrostatics show the uncertainty is significantly smaller than the underlying features in $\Delta G$ per particle using configuration mapping and SIMR.}
\end{figure}
\end{center}
\begin{center}
\begin{figure}[h]
\centering
{\centerline{\includegraphics[width=0.5\textwidth]{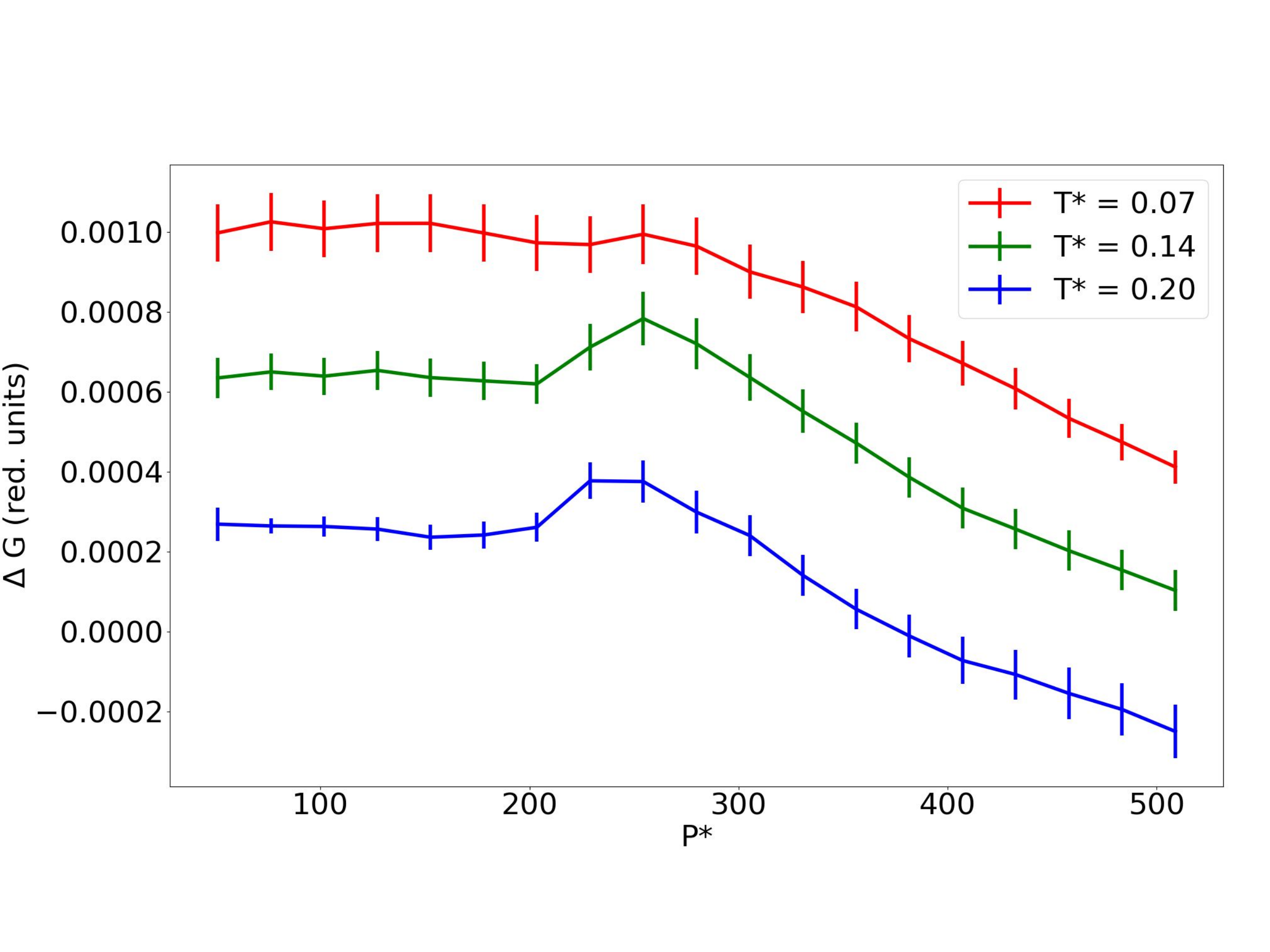}}}
        \caption{\label{fig:dgvp} The FCC-HCP free energy difference per particle and the associated uncertainty as a function of $P^{*}$ for the calculations using a PME treatment for the long-range electrostatics, obtained using configuration mapping and SIMR. The graphs show the reentrant behavior near $P^*=0.250$ as well as the significantly higher precision needed for these calculations than for the cutoff-based simulations.
}
\end{figure}
\end{center}
 
\section{Discussion and Conclusions}

This configuration mapping approach is likely to be useful for systems
of point particles, whatever the potential may be, as the same mapping
presented here will be valid.  Based on previous results with
configuration mapping on rigid water molecules in liquid
water~\cite{Paliwal2013}, this approach is likely to work with small rigid
molecules as well, but it is not clear how well it would work for more
complex molecules with internal degrees of freedom that will change
configuration ensemble between phases. There are a number of
improvements that can be made as well.  A number of additional
extensions may be possible. For example, adaptive choices of the
simulation points, as discussed in Schieber et
al.~\cite{Schieber2018}, can further decrease the clock time required
at the expense of wall time. Many of the state-to-state mappings still
lead to essentially negligible overlap, and hence may be unnecessary;
it may be possible to \textit{not} map those states together, saving some
amount of time spent mapping while losing negligible efficiency,
though determining exactly which states to exclude or include is
somewhat complicated.

The phase diagram of the solid phases of Lennard-Jones spheres has been predicted many times in literature, with large discrepancies between methods. The Successive Interpolation of Multistate Reweighting method (SIMR) is another method that can be used to determine the coexistence line of solid-solid transitions using full molecular dynamics simulations. However, this method is dependent on phase space overlap between adjacent simulations. Configuration mapping is a way to increase phase space overlap, and therefore computational efficiency in phase diagram prediction. 

Using a potential switch van der Waals cutoff, the coexistence curves were efficiently generated using this method to reasonable precision compared to the scale of $\Delta \mu$ between phases, and demonstrates the efficiency of SIMR plus configuration mapping in a standard problem for point particles, such as found in simulations of metals or inorganic materials. However, this cutoff approach introduces nonphysical behavior in the energy difference versus pressure curves, which can be understood by examining the radial distribution functions. As the pressure increases more layers of atoms, and thus peaks in the RDF, are brought under the value of the cutoff, which nonphysically affects the Gibbs free energy difference and thus the predicted coexistence line.  Using particle mesh Ewald to calculate long range interactions avoids nonphysical behavior in the energy differences between polymorphs, without extremely long cutoffs. The extremely small difference in potential energy and Gibbs free energy between phases and small difference in slope between the free energy surfaces makes the determination of this more accurate coexistence line challenging. In particular, at low pressures, the increased movement of the atoms decreases the effectiveness of the mapping, making overlap poor.  Current limitations in analyzing larger numbers of states with MBAR prevents fully characterizing the phase diagram at lower pressures, though the uncertainty and bias are already very low (at or below 0.001 E*). 

When using particle mesh Ewald, the phase diagram produced by the addition of configuration mapping to the SIMR method shows independence of cutoff and, despite the issues with statistical convergence, consistent trends in agreement with the current somewhat diverging literature results. The HCP phase is most stable at moderate pressures and low temperatures, with the FCC phase being more stable at high temperatures and extreme pressures. The coexistence curve displays reentrant behavior consistent with previous results. The maximum temperature of HCP stability is lower than the most likely most accurate results of Schultz et al.~\cite{Schultz2016}, likely due to bias in the PSCP value used to generate our coexistence curve. We found that a change in the reference value on the order of 0.0007 $E^*$ brings out curve to within uncertainty of these literature results. These results demonstrate that the method, although not perfect for the present calculation, should be effective for most problems of applicable interest.

\section{Supplementary Material}
Supplementary material online includes a discussion of the changes in
the free energy changes due to cutoffs, including an analysis of
changing radial distribution function changes as function of pressure,
as well as comparison between the uncertainty lines determined by
bootstraps over phase diagram lines and propagation of error in free
energy perpendicular to the tangent line.  We also include the GROMACS
input files used for simulations of Lennard-Jones particles for
reference.

\section{Acknowledgments}
This work used the Extreme Science and Engineering Discovery Environment (XSEDE), which is supported by National Science Foundation grant number OCI-1053575.  Specifically, it used the Bridges system, which is supported by NSF award number ACI-1445606, at the Pittsburgh Supercomputing Center (PSC). This work was supported financially by NSF through the grants NSF-CBET 1351635 and NSF-DGE 1144083. We thank Nate Abraham (CU Boulder) for discussions and comparisons, and Andrew Schultz (SUNY Buffalo) for helpful assistance in comparison to the results in Schultz et al.

\bibliography{bib}

\newpage
\input{supporting_short}

\end{document}

%% file: supporting_short.tex
\clearpage

\onecolumngrid
\begin{center}
\large{\bf{Supplementary Material: Configurational Mapping Significantly Increases the Efficiency of Solid-Solid Phase Coexistence Calculations via Molecular Dynamics: Determining the FCC-HCP Coexistence Line of Lennard-Jones Particles}}

Natalie P. Schieber, Department of Chemical and Biological Engineering, University of Colorado Boulder, Boulder, CO 80309, USA \\
Michael R. Shirts, Department of Chemical and Biological Engineering, University of Colorado Boulder, Boulder, CO 80309, USA \\
\end{center}

\twocolumngrid

\section{Nonphysical differences in the potential energy between phases as a function of pressure}\label{section:cutofferror}

\begin{figure}
\centering
\includegraphics[width=0.8\columnwidth]{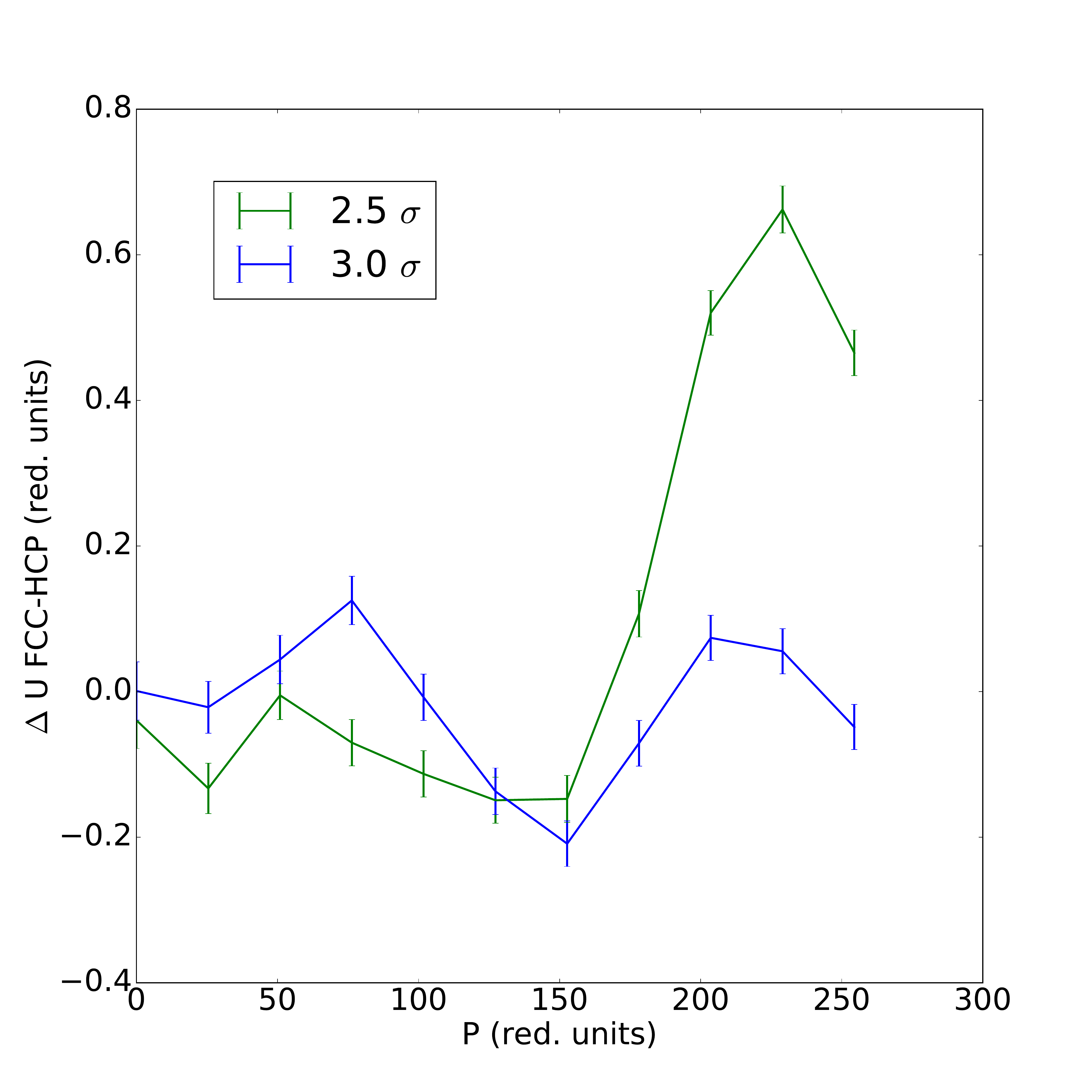}
        \caption{\label{fig:dUbaro} The difference between the energy of FCC and HCP phases for the $3.0 \sigma $ and $2.5 \sigma$ cutoff as a function of pressure shows non-monotonic behavior at T* = 0.47.} 
\end{figure}

The incorrect high pressure behavior in the phase diagram using a potential-switch cutoff is due to non-monotonic behavior in the difference in energy between polymorphs as a function of pressure, as seen in Figure~\ref{fig:dUbaro}. The `bumps' in the $\Delta U_{FCC-HCP}$ are present with both the $2.5 \sigma$ and $3.0 \sigma$ cutoffs and are well outside statistical uncertainty compared to the PME results. For these validation simulations, we use a Berendsen barostat~\cite{Berendsen1984} and velocity rescale thermostat of Bussi et al.~\cite{Bussi2007} in order to reduce statistical error with respect to a MTTK barostat and Nos\'{e}-Hoover thermostat used in other parts of the simulation. In all cases, MTTK and Berendsen ensemble averages were within uncertainty of each other as expected for equilibrium averages. These simulations were equilibrated for 14,849 $t^*$ (5 ns) and then ran for 14,849 $t^*$ (5 ns) with a 0.00297 $t^*$ (4 fs) time step to ensure sufficient equilibration. The magnitude of the `bumps' in the potential energy difference as a function of pressure decrease with increasing temperature, as would be expected as the simulation gets less ordered in the region of the cutoffs.

\begin{figure*}
\centering
\subfloat[]{\includegraphics[width=0.48\textwidth]{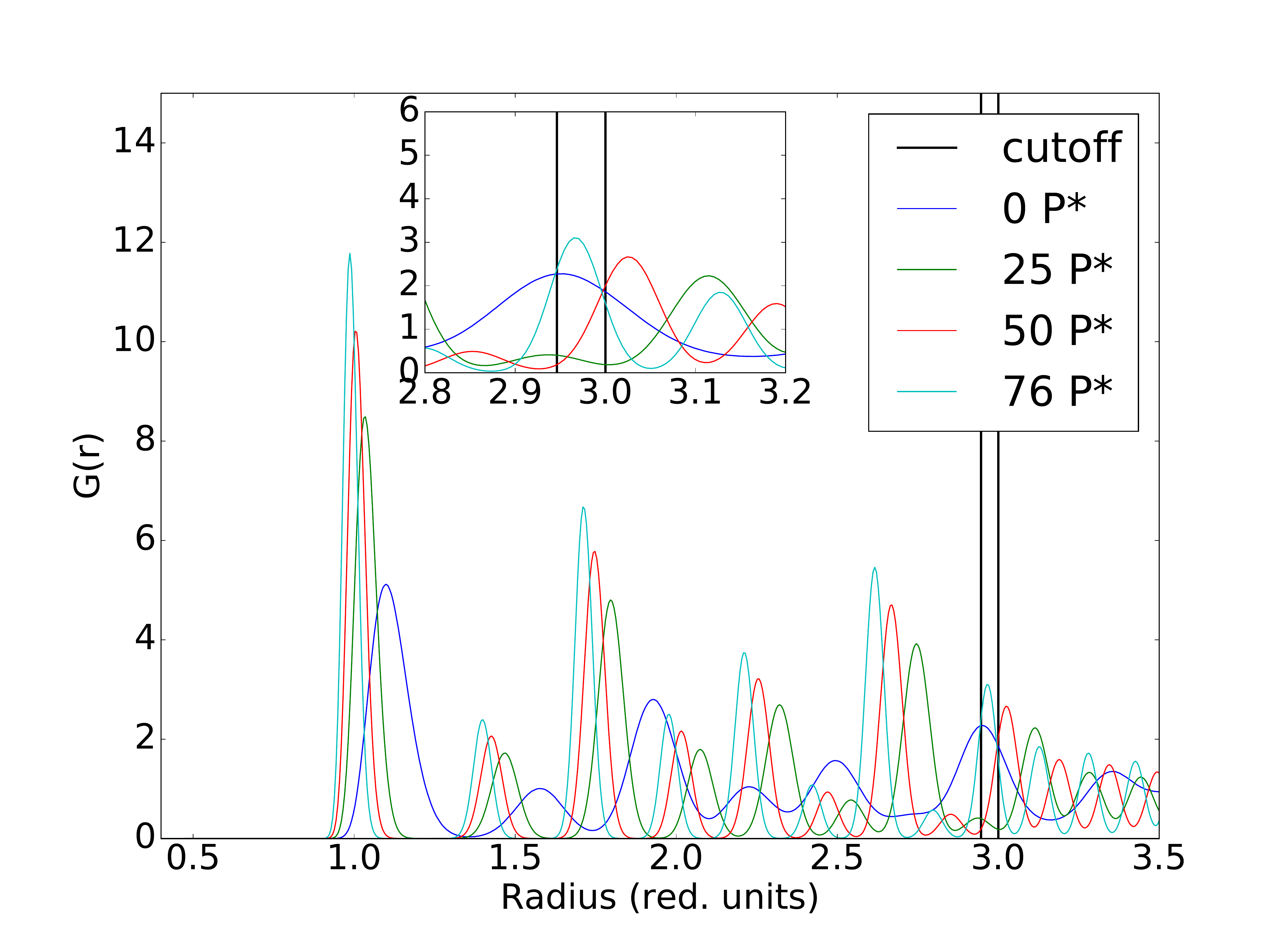}\label{subfig:fccrdf3}}%
\subfloat[]{\includegraphics[width=0.48\textwidth]{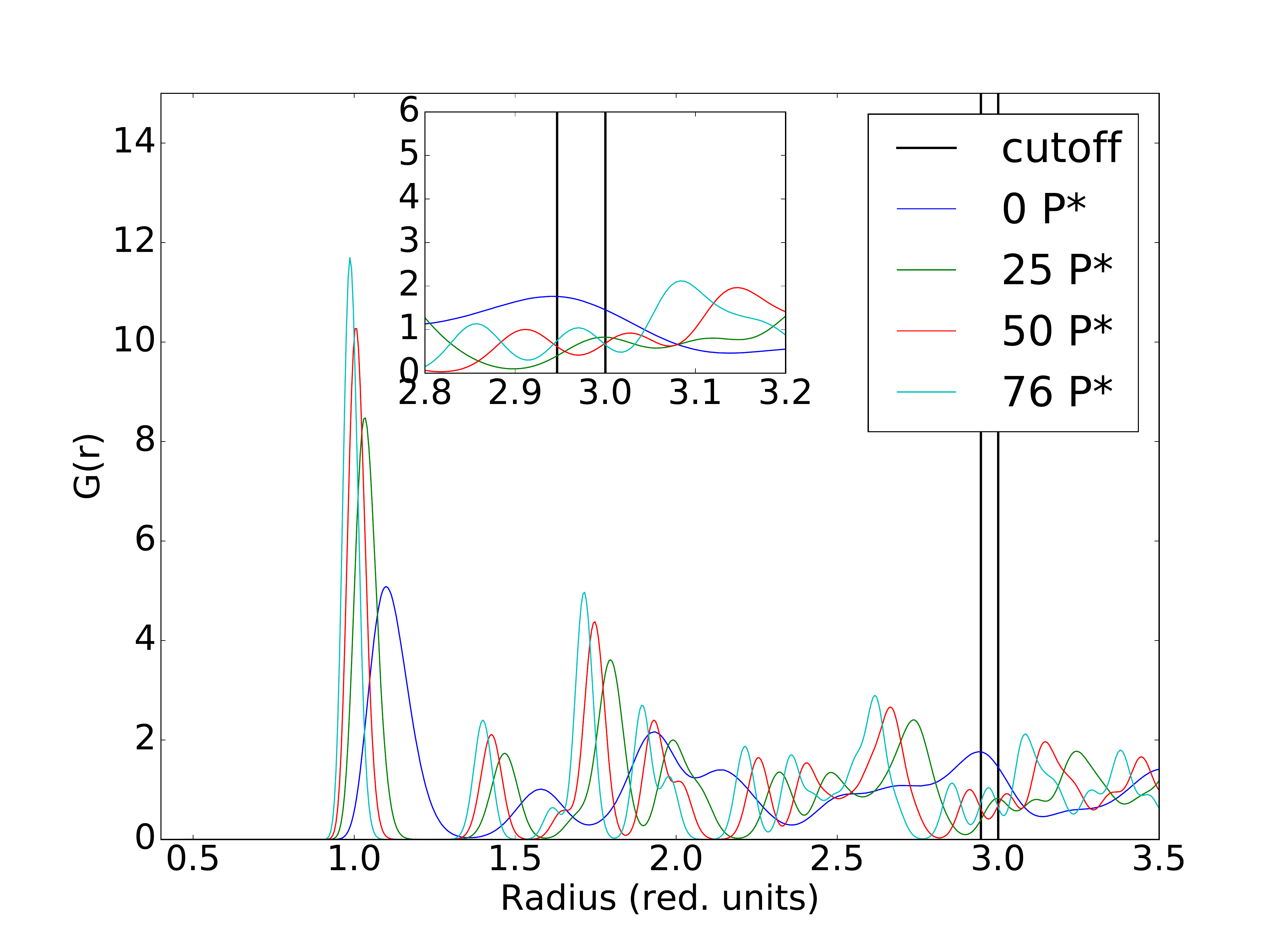}\label{subfig:hcprdf3}}\\
\subfloat[]{\includegraphics[width=0.48\textwidth]{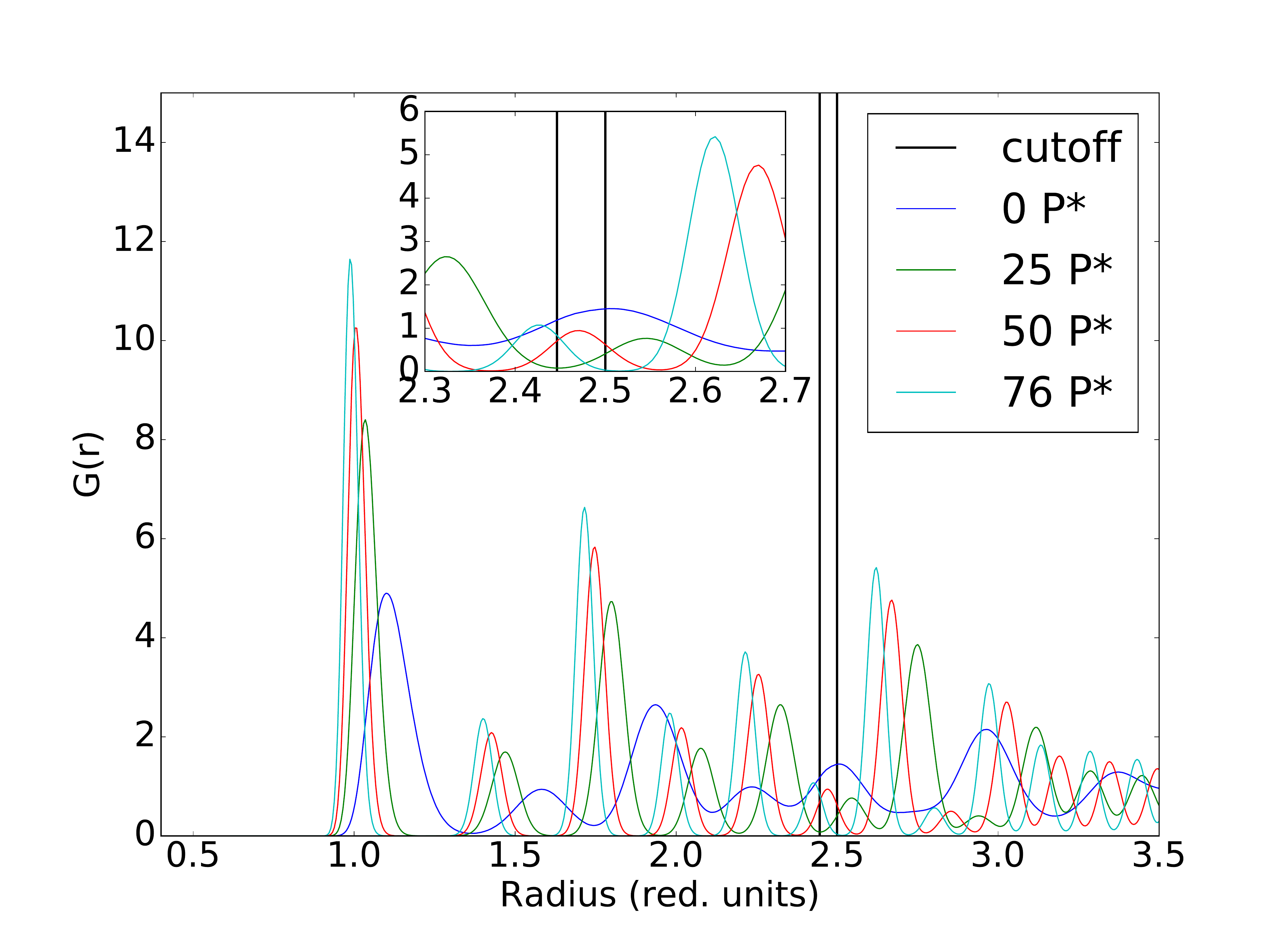}\label{subfig:fccrdf2}}%
\subfloat[]{\includegraphics[width=0.48\textwidth]{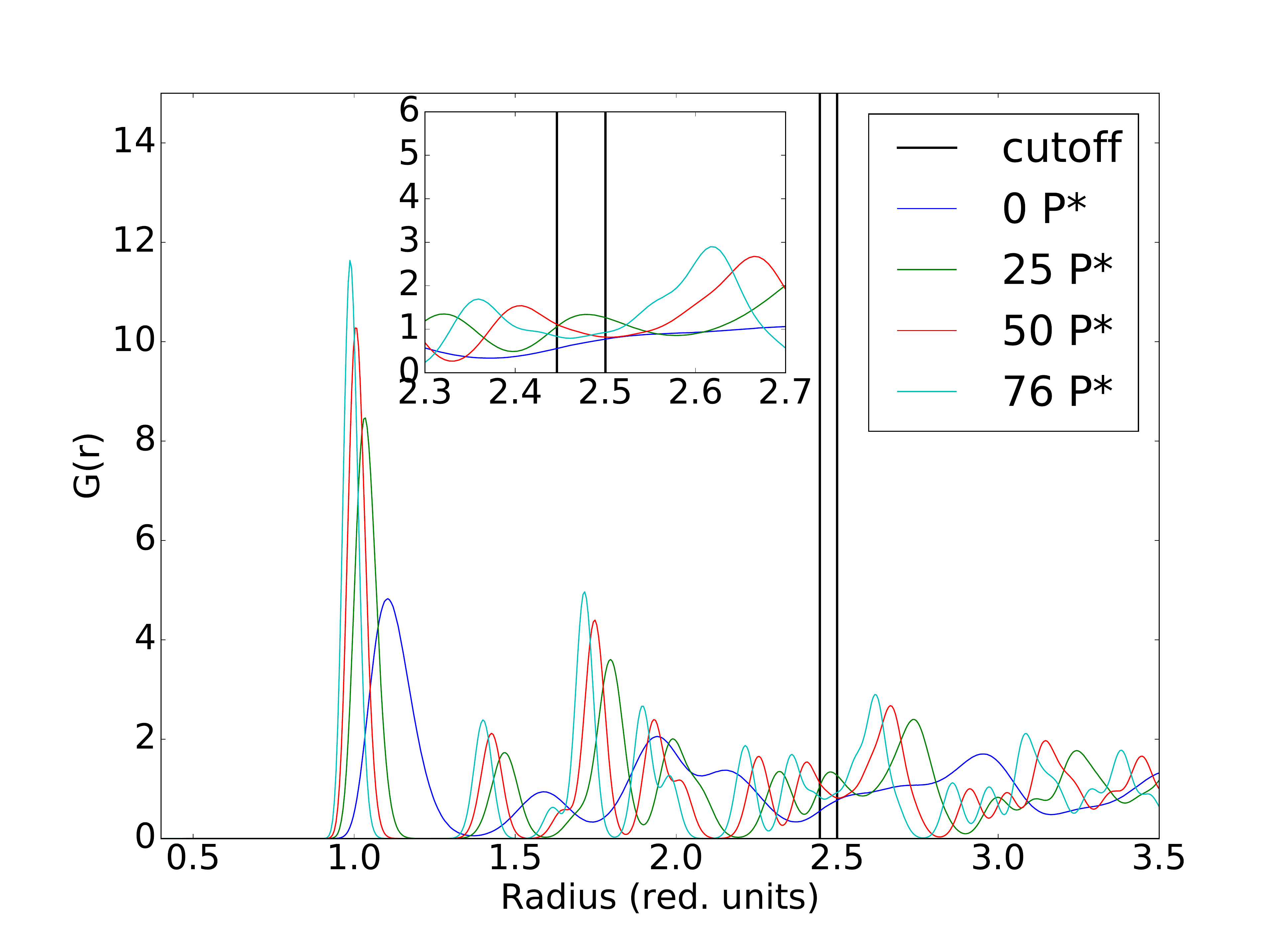}\label{subfig:hcprdf2}}
        \caption{\label{fig:rdfs} The radial distribution functions of the $3.0 \sigma$ cutoff simulations in FCC (\ref{subfig:fccrdf3}) and HCP (\ref{subfig:hcprdf3}) phases and $2.5 \sigma$ cutoff simulations in FCC (\ref{subfig:fccrdf2}) and HCP (\ref{subfig:hcprdf2}) phases show peaks jumping inside the cutoff (depicted as two lines, since it is a switched cutoff) as pressure increases.}
\end{figure*}

We hypothesize that these `bumps' in the $\Delta U $ versus pressure graphs are caused by a combination of the compressibility, ordered packing, and cutoff method used in this system. As the system compresses at higher pressures, a new layer of atoms around each atom of interest quickly `jump' into the radius under the cutoff. These abrupt jumps cause the non-monotonic behavior seen in Figure~\ref{fig:dUbaro}.  This behavior is seen in the radial distribution function (RDF) of the phases as a function of pressure in Figure~\ref{fig:rdfs}. The Berendsen barostat was again used for the radial distribution function comparison because it results in smaller fluctuations in the potential energy and volume and therefore a lower uncertainty. The number of peaks under the lower cutoff bound as a function of pressure for selected pressures is shown in Table~\ref{table:number}. For example, for the FCC structure, using the $2.5 \sigma$ cutoff, two additional peaks move into the cutoff bound between 0 $P^*$ and 250 $P^*$, while for the $3.0 \sigma$ cutoff, four additional peaks move into the cutoff. This is consistent with more bumps in Figure~\ref{fig:dUbaro} being seen in for the $3.0 \sigma$ results. 

\begin{table} \label{table:numbering}
\centering
\begin{tabular}{|ccccc|}
\hline 
 Pressure  $P^*$     & FCC $2.5 \sigma$  & FCC $3.0 \sigma$   & HCP $2.5 \sigma$  & HCP $3.0 \sigma$     \\
\hline
0.0025 &     4.5  & 5.5  &   4.5  & 6.0 \\ 
25 & 5.5  & 7.5  &   6.5  & 8.5  \\ 
50 & 6.0  & 8.0 &   7.75  & 9.75  \\
76 & 6.0  & 8.5  &   8.0  & 10.5  \\
127 & 6.0  & 9.0  &   9.0  & 13.0  \\
254 & 6.5 & 9.5 &   11.0  & 15.0  \\
\hline
\end{tabular}
\caption{\label{table:number} The number of peaks below the cutoff for the FCC and HCP phases as a function of pressure.}
\end{table}

\begin{center}
\begin{figure}
\includegraphics[width=0.8\columnwidth]{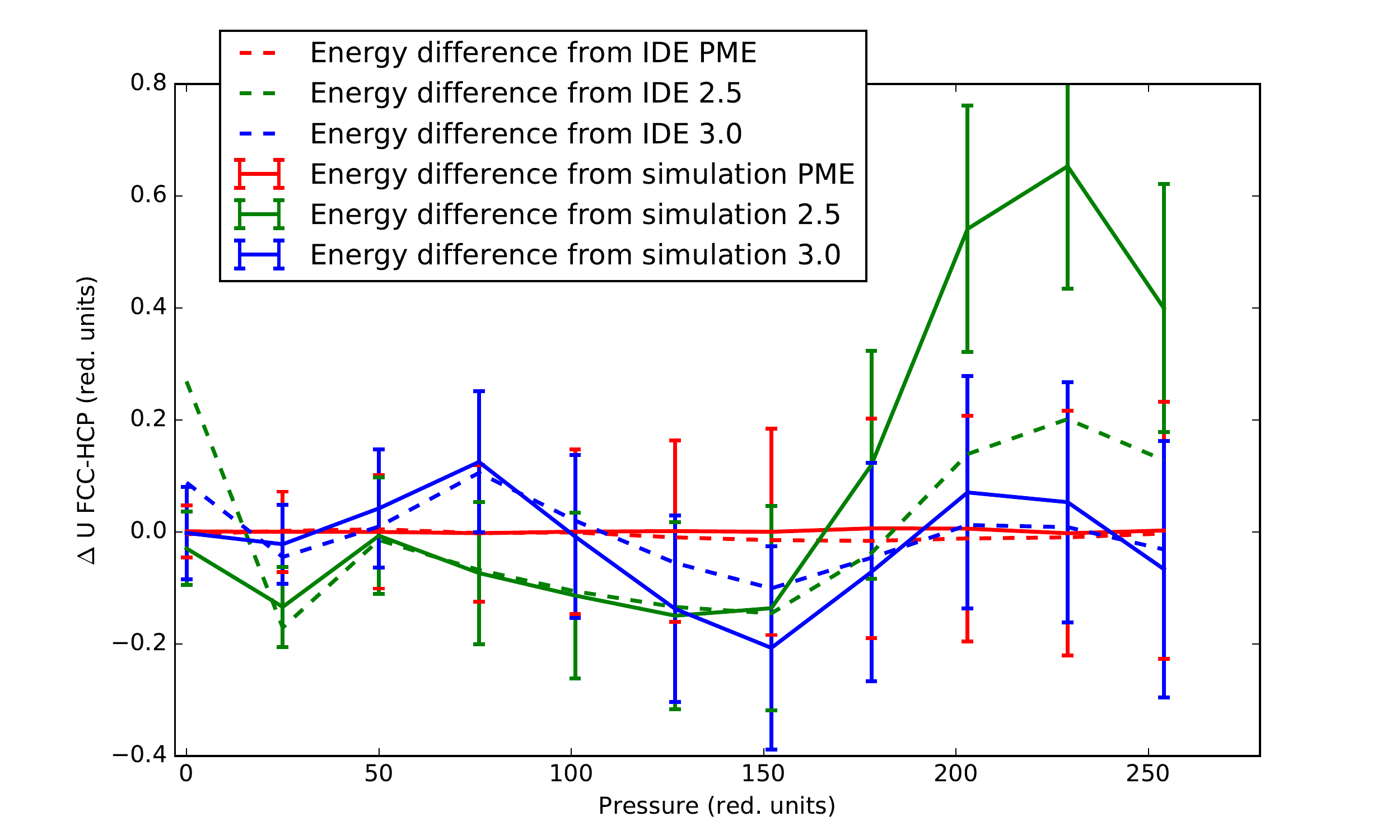}
\caption{\label{fig:rdfint} The difference in the integrated radial distribution functions of the FCC and HCP phases at cutoffs of $2.5 \sigma$ and $3.0 \sigma$ as well as using PME show the same non-monotonic behavior as the $\Delta U$ versus pressure curves for the simulations treated with a potential switch cutoff and flat behavior for PME simulations.}
\end{figure}
\end{center}

The effect that the movement of RDF peaks through the cutoff has on the energy and volume difference between phases is verified by integrating the contribution of the dispersion energy under the cutoff as a function of pressure. To estimate this difference, we determined the radial distribution function at each pressure and then integrated the Lennard-Jones dispersion energy over this particle-particle density to obtain $\int_{0}^{r_c} -4\epsilon r^{-6} g(r) 4\pi r^{2} dr$, where $r_c$ is the cutoff distance. The difference in this integral between phases is shown in Figure~\ref{fig:rdfint} and shows the same qualitative behavior as a function of pressure as seen in Figure~\ref{fig:dUbaro} for both cutoff distances tested. The main source of error in the $\Delta U$ curves is thus due to the uneven incorporation of shells of atoms into the RDF resulting from the use of a cutoff. In contrast, the directly calculated energy difference and dispersion energy difference using PME are almost independent of pressure, as seen in Figure~\ref{fig:rdfint}. 

\section{Differences in error estimation methods}\label{section:errortype}

\begin{figure}[H]
\begin{center}
\subfloat[]{\includegraphics[width=0.9\columnwidth]{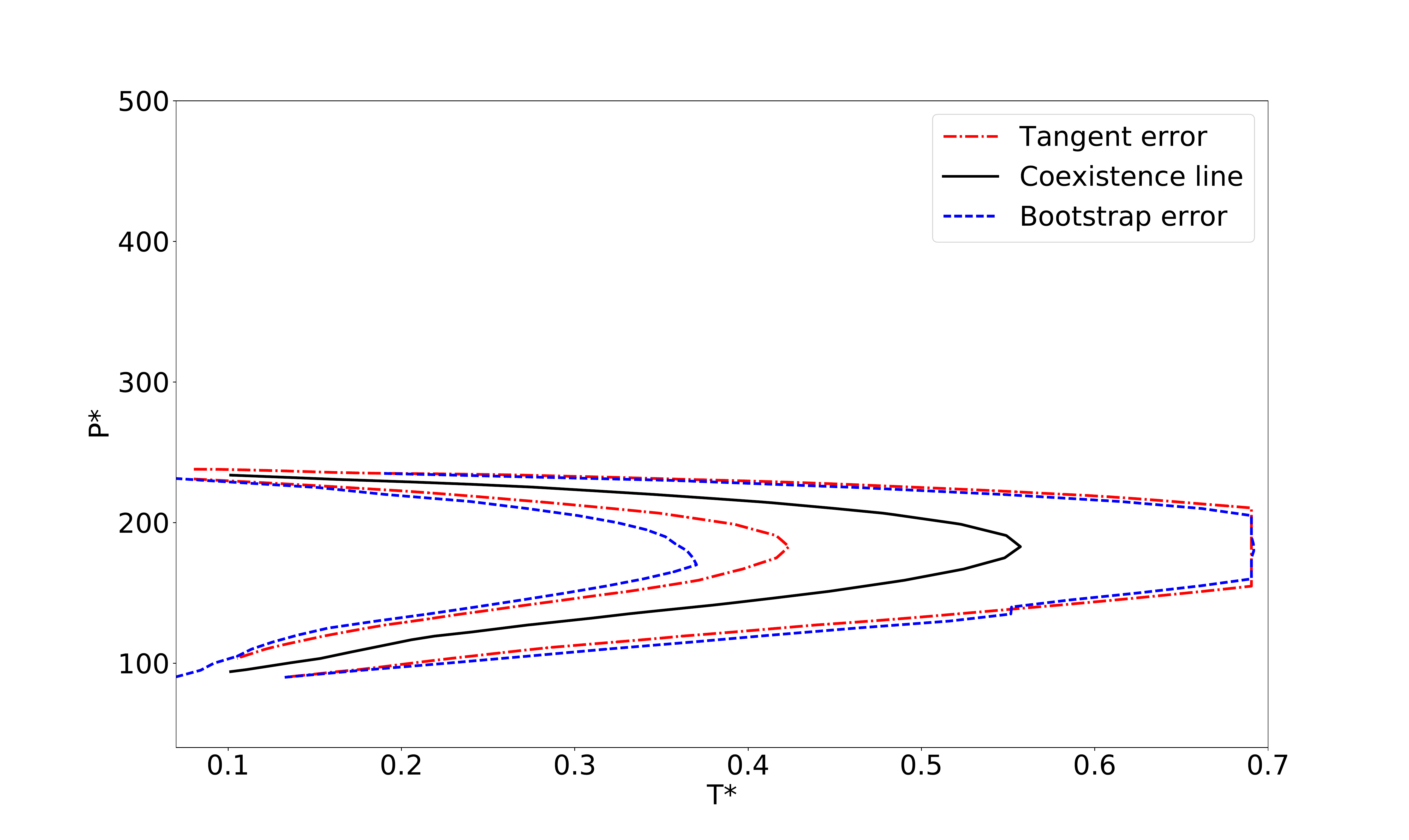}\label{subfig:bstrap25}}\\
\subfloat[]{\includegraphics[width=0.9\columnwidth]{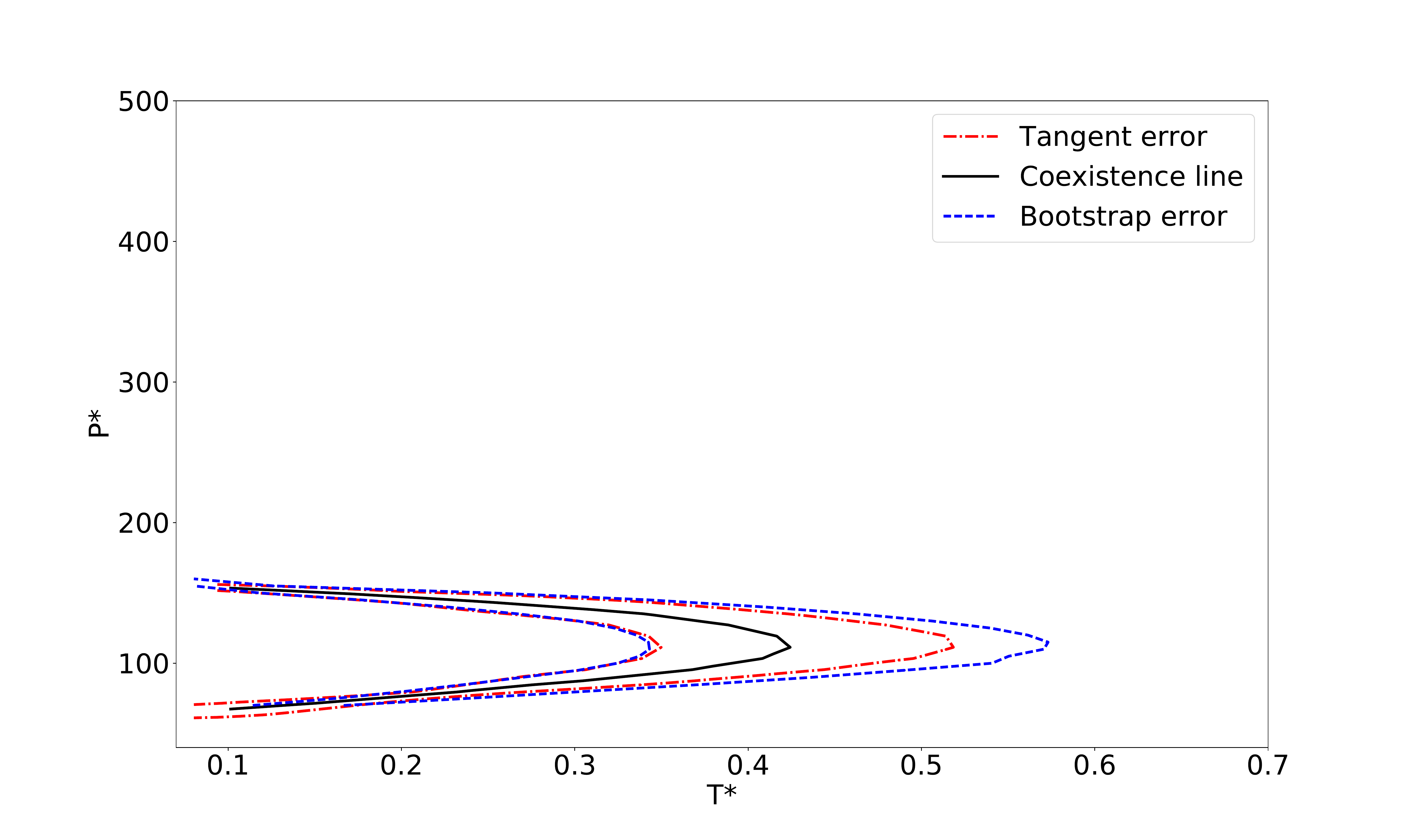}\label{subfig:bstrap30}}\\
        \caption{\label{fig:bstrapcompare} The coexistence lines with two methods of error estimation for a potential shift cutoff of $2.5 \sigma$ (\ref{subfig:bstrap25}) $3.0 \sigma$ (\ref{subfig:bstrap30}) shows consistent behavior between the two error methods.}
\end{center}
\end{figure}

\begin{figure}
\centering
{\includegraphics[width=0.9\columnwidth]{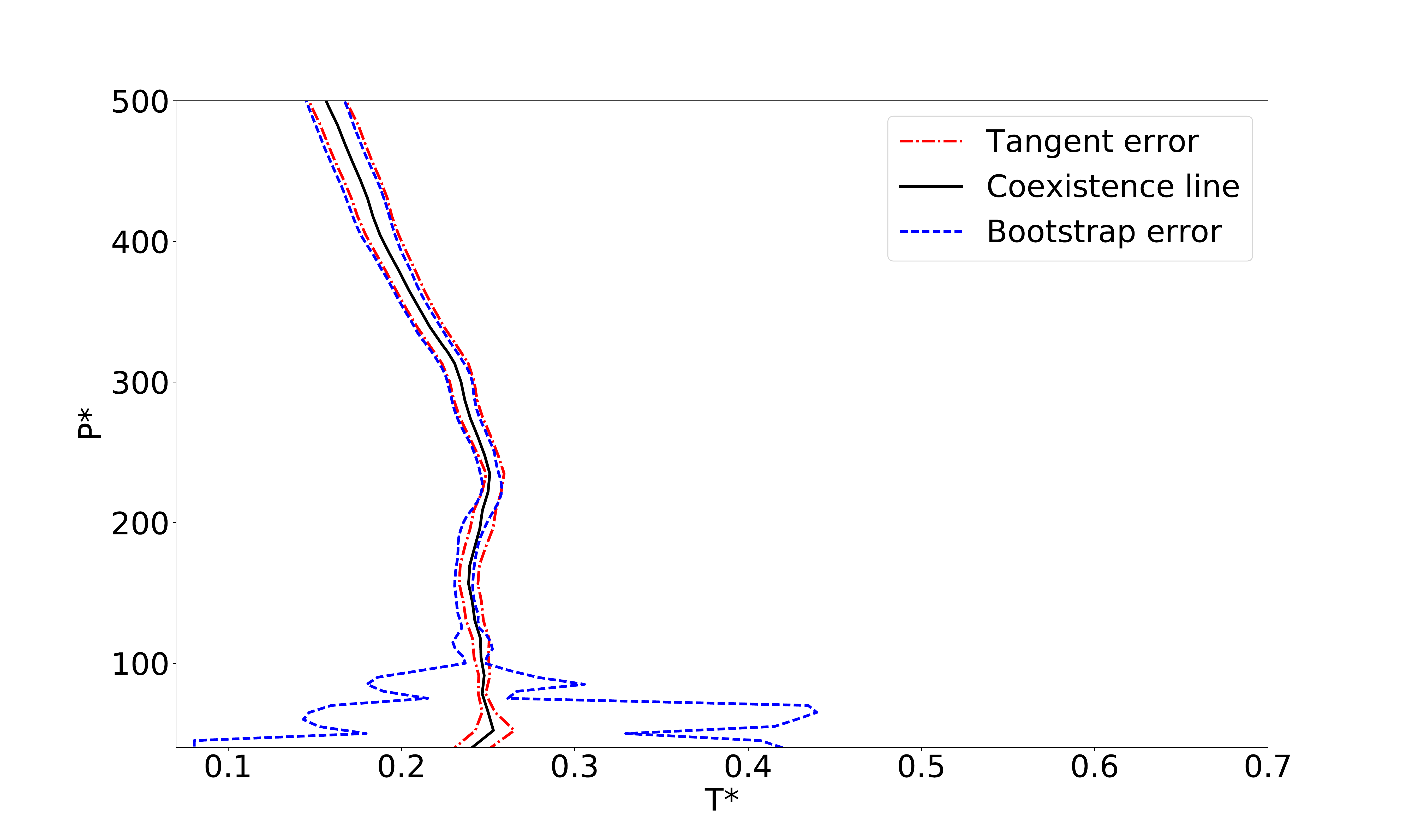}\label{subfig:bstrapPME}}\\
        \caption{\label{fig:bstrapcomparePME} The coexistence lines with two methods of error estimation  with the PME treatment of long range interactions shows consistent behavior between the two error methods at high pressure and poor agreement at low pressure.}
\end{figure}

When estimating the uncertainty in the coexistence line, we have tested two ways of plotting the error. In the first method, the uncertainty in the $\Delta G_{FCC-HCP}$ value is calculated, either by bootstrapping the configurations used in the MBAR calculation or by relying on the uncertainty estimation built into \texttt{pymbar}. This is then transformed into the uncertainty in the actual width of the coexistence line using the method described in \cite{Schieber2018}. Alternately, the entire process can be bootstrapped. In this method the configurations used in the MBAR calculations are bootstrapped and a coexistence line is generated for each set of configurations, resulting in a set of $\sim200$ lines, with possibly fewer if some fail to converge due to poor overlap in the free energy calculation. To find the uncertainty bounds on the coexistence line, the set of lines are arranged from smallest to largest from the median in each direction. The line representing the desired confidence interval (in the case of these calculations, one standard deviation) from the median is determined in each direction and is the uncertainty bound in that direction. For the case of HCP and FCC spheres, the temperature dimension was used as the axes that the lines were arranged along, though with less noisy data, one could do this calculation in the tangent direction. A comparison of the two error methods can be seen for the three coexistence lines can be seen in Figs.~\ref{fig:bstrapcompare} and ~\ref{fig:bstrapcomparePME}. 

For the two uncertainty lines that results from simulations with a potential switch cutoff, the two error plotting methods are consistent. They agree to within 20\% at all points. For the systems studied in this case, SIMR with mapping produced good quality coexistence lines. However, the coexistence line from the PME simulations has error estimations that diverge significantly at low pressures. The set of 200 bootstrap lines diverges greatly into two directions at this point. This means that while the standard deviation of the $\Delta G $ value is reasonable, the actual lines themselves are very different. This is due to poor overlap at low pressures. The amount of movement in the atoms causes the mapping of the pressures to be less effective, and this combined with the extremely small free energy differences in this region, means that the phase diagram cannot be easily and reliably determined for low pressures with this method.
